\documentclass[prl,twocolumn,superscriptaddress]{revtex4-2}   
\usepackage{graphicx,amssymb}
\usepackage{color,ulem}
\usepackage{bm}   
\usepackage{lineno}
\usepackage{bm}
\usepackage{amssymb,amsmath}
\usepackage{amsthm}
\usepackage{latexsym}
\usepackage{graphicx,amssymb}
\usepackage{color,ulem}
\usepackage{float}
\usepackage{siunitx}
\usepackage{mathrsfs} 
\usepackage{soul}
 
\usepackage{graphicx}
\usepackage{dcolumn}
\usepackage{bm}
\usepackage[utf8]{inputenc}
\usepackage{amsfonts,amssymb,amsmath}
\usepackage[mathscr]{eucal}
\usepackage{epsfig}

\usepackage{rotating}

\begin{document}
	
	\title{Quantum-Ready Microwave Detection with Scalable Graphene Bolometers in the Strong Localization Regime}
	
\author{Yu-Cheng Chang}
\affiliation{Pico group, QTF Centre of Excellence, Department of Applied Physics, Aalto University, P.O. Box 15100, FI-00076 Aalto, Finland}
\author{Federico Chianese}
\affiliation{Department of Microtechnology and Nanoscience, Chalmers University of Technology, 412 96 Gothenburg, Sweden}
\author{Naveen Shetty}
\affiliation{Department of Microtechnology and Nanoscience, Chalmers University of Technology, 412 96 Gothenburg, Sweden}
\author{Johanna Huhtasaari}
\affiliation{Department of Microtechnology and Nanoscience, Chalmers University of Technology, 412 96 Gothenburg, Sweden}
\author{Aditya Jayaraman}
\affiliation{Department of Microtechnology and Nanoscience, Chalmers University of Technology, 412 96 Gothenburg, Sweden}
\author{Joonas T. Peltonen}
\affiliation{Pico group, QTF Centre of Excellence, Department of Applied Physics, Aalto University, P.O. Box 15100, FI-00076 Aalto, Finland}
\author{Samuel Lara-Avila}
\affiliation{Department of Microtechnology and Nanoscience, Chalmers University of Technology, 412 96 Gothenburg, Sweden}
\author{Bayan Karimi}
\affiliation{Pico group, QTF Centre of Excellence, Department of Applied Physics, Aalto University, P.O. Box 15100, FI-00076 Aalto, Finland}
\affiliation{Pritzker School of Molecular Engineering, University of Chicago, Chicago IL 60637, USA}
\author{Andrey Danilov}
\affiliation{Department of Microtechnology and Nanoscience, Chalmers University of Technology, 412 96 Gothenburg, Sweden}
\author{Jukka P. Pekola}
\affiliation{Pico group, QTF Centre of Excellence, Department of Applied Physics, Aalto University, P.O. Box 15100, FI-00076 Aalto, Finland}
\author{Sergey Kubatkin}
\email{sergey.kubatkin@chalmers.se}
\affiliation{Department of Microtechnology and Nanoscience, Chalmers University of Technology, 412 96 Gothenburg, Sweden}
\affiliation{InstituteQ – the Finnish Quantum Institute, Aalto University, Finland}

	\date{\today}
	
	\begin{abstract}
		Exploiting quantum interference of charge carriers, epitaxial graphene grown on silicon carbide emerges as a game-changing platform for ultra-sensitive bolometric sensing, featuring an intrinsic resistive thermometer response unmatched by any other graphene variant. By achieving low and uniform carrier densities, we have accessed a new regime of strong charge localization that dramatically reduces thermal conductance, significantly enhancing bolometer performance.  Here we present scalable graphene-based bolometers engineered for detecting GHz-range photons, a frequency domain essential for superconducting quantum processors. Our devices deliver a state-of-the-art noise equivalent power of 40 zW$/\sqrt{\rm Hz}$  at $T=40$~mK, enabled by the steep temperature dependence of thermal conductance, $G_{\rm th}\sim T^4$ for $T<100~$mK. These results establish epitaxial graphene bolometers as versatile and low-back-action detectors, unlocking new possibilities for next-generation quantum processors and pioneering investigations into the thermodynamics and thermalization pathways of strongly entangled quantum systems.
	\end{abstract}
	\maketitle

	Detecting extremely small powers of electromagnetic radiation is essential in modern physics, where even the faintest signals can carry valuable information. 
	One prominent example is the search for axions—hypothetical particles that could make up dark matter—which undergo conversion into photons in a magnetic field, thus producing an extremely weak microwave radiation~\cite{adair_search_2022,braine_extended_2020,pankratov_detection_2025}. 
	Another example is quantum information processing, where operations on qubits often involve single or few-photon signals that require precise, low-noise detection. 
	In both cases, bolometers provide an effective solution~\cite{irastorza_new_2018,sikivie_experimental_1983,shokair_future_2014,karimi_quantum_2020,gunyho_single-shot_2024}. These devices are temperature-dependent resistors that absorb radiation and are weakly thermally coupled to a reservoir, allowing them to register minute temperature increases. This enables accurate detection of tiny power levels without adding significant electrical noise. In a bolometer, minimizing electron-phonon coupling to the substrate is 
	crucial to ensure that absorbed radiation effectively heats the electron system without rapidly dissipating energy to the lattice.  
	
	Graphene doped to the Dirac point is an ideal material for sensitive detection because a shrinking Fermi surface at extremely low electron density imposes phase-space restrictions for electron-phonon interactions. 
	This allows graphene to retain absorbed energy longer, resulting in higher temperature sensitivity and improved performance for detecting extremely small power levels.
	However, at high carrier density graphene’s intrinsic temperature dependence of resistivity is too weak for efficient bolometer operation~\cite{mckitterick_ultrasensitive_2015,efetov_controlling_2010}. To address this limitation, one can induce superconductivity in graphene by coupling it to a nearby superconductor
	and leverage the superconducting transition. Indeed, impressive results were achieved on a Superconductor–Graphene hybrid junction, using an exfoliated graphene flake insulated by hexagonal boron nitride (hBN)~\cite{kokkoniemi_bolometer_2020}. Open issues that remain in this approach are scalability and, importantly, extending the operational temperature range of the proximitized-graphene bolometer, which is bound to operate close to the superconducting transition in the superconductor.
	
	\begin{figure*}
		\centering
		\includegraphics [width=\textwidth] {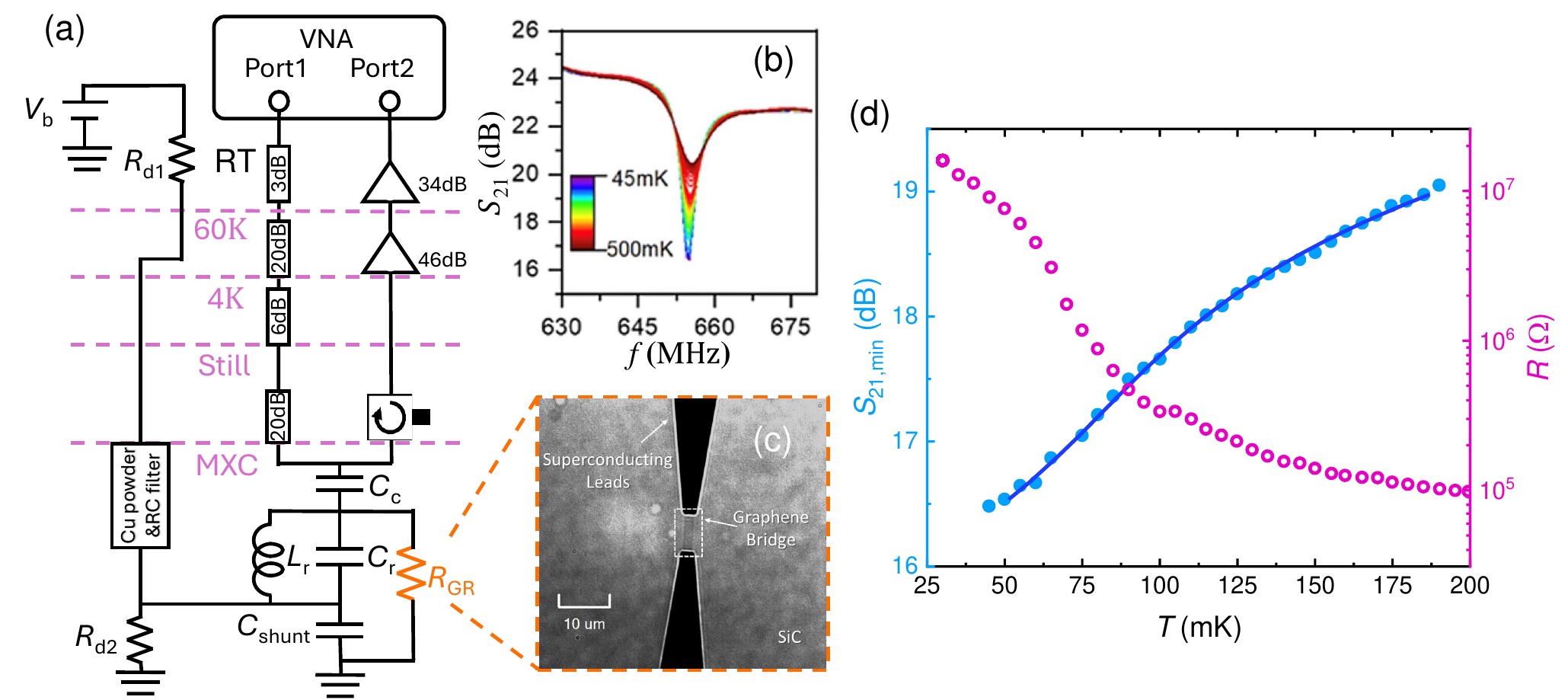}
		\caption{Monitoring graphene electron temperature by Radio Frequency (RF). (a) Cryogenic set up for RF notched transmission measurements on graphene bridge. (b) A set of resonances on $S_{21}$ transmission measured at different temperatures. The resonance quality factor follows the temperature dependence of graphene bridge resistance. (c) An optical transmission image of a $8\times 2~\mu{\rm m}^2$ graphene bridge, contacted by superconducting aluminum electrodes to eliminate the electronic heat transport to the leads; the $8~\mu{\rm m}$ long graphene bridge suppresses undesired supercurrent due to proximity effect. (d) Transmission $S_{21}$ for monitoring the graphene resistance and electronic temperature. The minima on transmission $S_{21} (T)$ at resonance (cf Fig.~\ref{fig1}b) vs. the fridge temperature $T$ (blue plot) and Direct Current (DC) resistance measurements $R(T)$ made on the same bridge (red plot). Together, $R(T)$ and $S_{21} (T)$ establish one-to-one $S_{21} (T)\leftrightarrow R$ correspondence. Both $R(T)$ and $S_{21} (T)$ have a tendency to saturate at lower temperatures – possibly due to improper thermalization (cf main text).
			\label{fig1}}
	\end{figure*}

	Here, we overcome these two critical shortcomings by introducing a novel platform based on epitaxial graphene grown on silicon carbide (SiC), a wafer-scale technology, allowing uniform graphene doping close to Dirac point~\cite{he_accurate_2022}. In this system the graphene layer exhibits an intrinsic temperature dependence of resistance caused by quantum interference of carriers in the presence of strong inter-valley scattering~\cite{mccann_weak-localization_2006,lara-avila_disordered_2011,ponomarenko_tunable_2011} - a mechanism fundamentally different from superconducting proximity effects. Crucially, this response is not bound by a superconducting critical temperature; instead, device sensitivity improves continuously as temperature decreases, making it exceptionally well-suited as sensitive cryogenic radiation detector. We demonstrate that below 200 mK, quantum interference effects in graphene induce strong localization of charge carriers, pushing the graphene bolometer into a qualitatively new transport regime. This leads to a remarkable enhancement in both responsivity and thermal isolation from the reservoir, with the potential to enable unprecedented sensitivity of truly scalable graphene detectors. 
	
	Importantly, the presented  graphene sensor has an extremely low heat capacity approaching that of a single degree of freedom, which translates into projected energy resolution in calorimeter mode, envisioned in~\cite{roukes_yoctocalorimetry_1999}, as low as $ k_BT \simeq 0.7$ yJ $=0.7 \times 10^{-24}$ J at 50 mK. Thus it would provide a click-detector resolving a single photon with frequency above 1 GHz, which is a component of the future quantum information processing toolbox that long has been desired. Thanks to drastically different physical mechanism behind bolometric operation as compared to previous reports of graphene-based detectors, our devices perform at the state-of-the-art level, making them promising for  applications in quantum information processing with superconducting qubits operating in the GHz range, and for fundamental studies in this field.
	
	\section*{Sample design and experimental setup}
	We have fabricated a wafer with 16 graphene strips with different aspect ratios, and three Hall bars, aimed to evaluate transport properties of graphene (The chip layout is in Supplementary Material, sec. S1). To thermally isolate graphene, the current leads were made of aluminum, a superconductor with $T_c$ of about $1~$K. (See Fig. S1 in Supplementary Information.) Hall bar measurements confirmed the targeted carrier density of $n\approx 10^{10}$~electrons$/{\rm cm}^2$, achieved by polymer-molecular doping~\cite{he_uniform_2018}. Magneto transport data of the longitudinal resistance hinted that localization in graphene is caused by quantum interference effects, which can be eliminated by applying a modest magnetic field, breaking time-reversal symmetry (see Fig. S2 in Supplementary Information). We conclude that upon lowering electron temperature the quantum-mechanical interference of the charge carriers in graphene increases, and sets graphene into the strong localization regime, where effects of electron-electron interactions are further contributing to a steeper $T$-dependence of $R$. 
	
	 In this report we will focus on the performance of two samples with higher aspect ratio, demonstrating best bolometric performance. They were 2 by 8 micrometers strips shown in Fig.~\ref{fig1}a, measured in two different dilution refrigerators with base temperatures down to 60 and 30 mK. Above 60 mK both samples demonstrated consistent results, even though they were evaluated in different measurement set-ups. In what follows, we analyze the data taken on the sample A measured down to the lowest temperature, and present the performance of the reference sample B in the heat conductance measurements (Fig. 3a). 
	
	\begin{figure*}
		\centering
		\includegraphics [width=\textwidth] {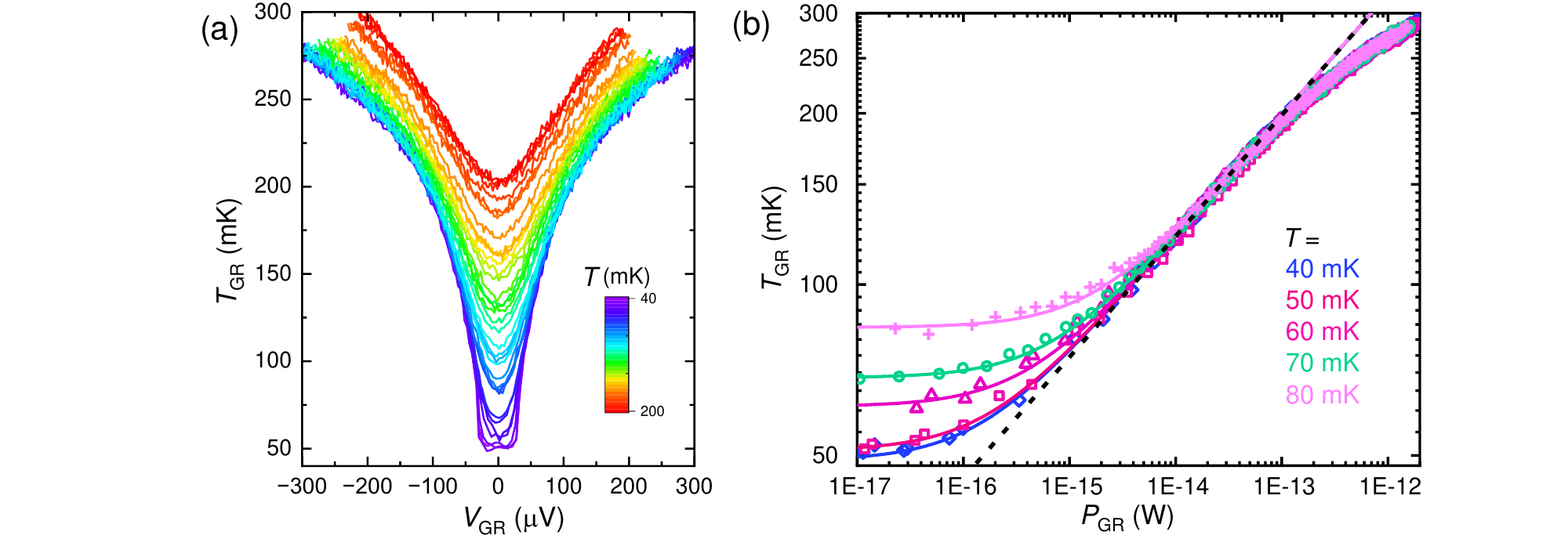}
		\caption{(a) Graphene electronic temperature $T_{\rm GR}$ as a function of the DC bias voltage $V_{\rm GR}$; a set of curves taken for different cryostat temperatures. (b) Same data presented as a function of the injected DC power $V_{\rm GR}^2/R (T_{\rm GR})$, only the data for the lowest fridge temperatures $45-80~$mK are presented for clarity. Symbols are for experimental data taken at cryostat temperatures 40~mK (blue rhombs), 50~mK (red squares), 60~mK (magenta triangles), 70~mK (green circles) and 80~mK (violet crosses). The solid curves are fits to a power law $T_{\rm GR}=(P_{\rm GR}/\sigma+T_0^\beta)^{(1/\beta)}$ with parameters $\beta=4.7,\sigma=2\times 10^{-10}~{\rm W}/{\rm K}^\beta$, and  $T_0=49.5,~51.0,~61.0,~68.5,~84.0$~mK. The straight black dashed line has a slope 4.7 $(P_{\rm GR}\sim T_{\rm GR}^{4.7})$.
			\label{fig2}}
	\end{figure*}
	The experimental measurement setup is schematically presented in Fig.~\ref{fig1}; essentially we measure the heat conductance between the graphene electronic system and the substrate, using quantum interference - induced $R(T)$ as a thermometer for graphene electronic system, assuming that the substrate temperature follows the base temperature of the cryostat. To monitor graphene resistance with good precision and minimal heat load we used an RF scheme, suggested in~\cite{schoelkopf_radio-frequency_1998}: relatively high graphene resistance was matched to the input resistance of a low noise amplifier at a frequency of 650 MHz by a lumped-element resonator with a quality factor $Q\approx 100$. A separate DC line allowed us to apply a control power to the graphene sample (See Fig.~\ref{fig1}a). The RF scheme was accompanied by a DC measurement setup, which was also used for the bolometer calibration with heating at a known Joule power, while monitoring the electronic temperature of graphene by the RF conductance measurements. To avoid sample overheating by the RF probe power, we performed measurements at low RF powers – the detailed information will be presented in the discussion section and Supporting Information. The DC bias line was filtered/thermalized by the termocoax, followed by lumped element filters - its thermalization was checked in separate experiments. 
	
	\begin{figure*}
		\centering
		\includegraphics [width=\textwidth] {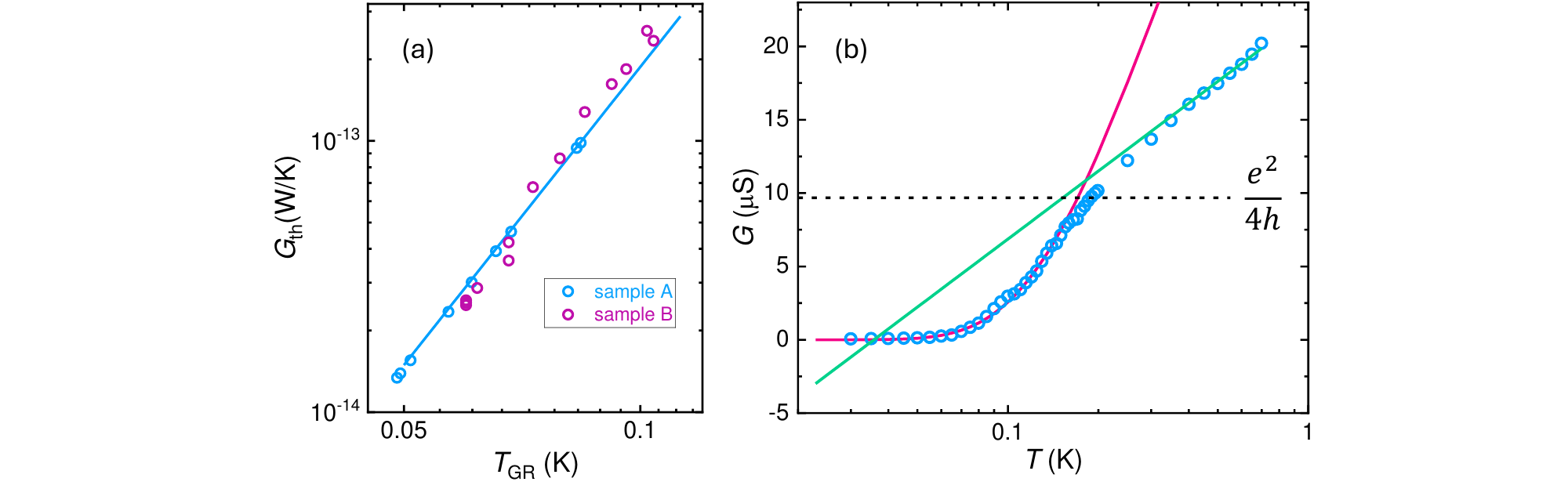}
		\caption{(a) Graphene thermal conductance $G_{\rm th} (T_{\rm GR})=\beta \sigma T_{\rm GR}^{(\beta-1)}$ as a function of graphene temperature $T_{\rm GR}$, 
			presented are the data for two samples with of identical dimensions $W\times L=2\times 8~\mu{\rm m}^2$, 
			measured in two different cryostats (with base temperatures 30~mK for sample A and 60~mK for sample B). Both data sets have similar slope $\beta -1\approx 3.7$ in log-log plot. 
			The blue solid line is for fit $G_{\rm th} (T_{\rm GR})=5\sigma T_{\rm GR}^{3.7}$ with $\sigma=2\times 10^{-10}~{\rm W}/{\rm K}^{3.7}$. 
			(b) Graphene conductance as a function of temperature. Red curve is a fit to Arrhenius law $\ln (G_{\rm GR})=-E_{\rm Arr}/(k_B T_{\rm GR})-9.67$ with $E_{\rm Arr}=k_B\times 0.32$~K. 
			Green line presents logarithmic dependence $\frac{G_\square}{4}\, \frac{e^2}{h} \ln(\frac{T}{T'})$ with $G_\square=0.69$ and $T'= 0.0358$~K.
			Transition from weak to strong localisation occurs at per square conductivity of the order of $~e^2/h$ (dashed black line), consistent with interference-induced Anderson localization 
			picture~\cite{ponomarenko_tunable_2011}.
			\label{fig3}}
	\end{figure*}
	Figure~\ref{fig1}d presents the calibration data for RF read out of the graphene electronic temperature: both graphene resistance $R$ and transmission resonance $S_{21,{\rm min}}$ were measured as a function of the fridge temperature $T$. Cooperatively, $R(T)$ and $S_{21,{\rm min}}(T)$ define the minimum on transmission $S_{21,{\rm min}}$ as a function of graphene resistance $S_{21,{\rm min}}(R)$ (technically, both $R(T)$ and $S_{21,{\rm min}}(T)$ were interpolated with polynoms - {\it cf } Supplementary Sec. 3,4 - to compose the $S_{21,{\rm min}}(T)$  functional dependence).
	
	\section*{Bolometric response of graphene sensor}
	Having the thermometry set up calibrated (Fig.~\ref{fig1}d), we applied a DC voltage to graphene terminals, overheating the strip, and measured the graphene electronic temperature in response to DC heating. The result is shown in Fig.~\ref{fig2}a, where we present graphene electronic temperature $T_{\rm GR} $ as a function of the DC bias voltage; the cryostat temperature  $T$ being the parameter. Note that as the bridge resistance diverges at $T\rightarrow 0$, for any given bias voltage $V$ the injected power $V^2/R\rightarrow 0$, so that below some temperature the power dissipated on graphene becomes lower than uncontrolled heat flux from non-thermalized environment (see sec. Discussion and Outlook for details). This is why the $T_{\rm GR}$ ($V_{\rm GR}$) curves taken at the lowest temperatures ($\leq 50$~mK) are flattened around the minima.
	
	As the next step, we can map the voltages on the horizontal axis onto dissipated  powers: for any point $(V_{\rm GR},T_{\rm GR})$ in Fig.~\ref{fig2}a the calibration curve in Fig.~\ref{fig1}d (blue) gives us the graphene temperature, and the corresponding dissipated power is thus $V_{\rm GR}^2/R(T_{\rm GR})$. The
	result of this transformation is presented in Fig.~\ref{fig2}b, only the plots for the lowest fridge temperatures are shown. Figure 2b is the central plot of this contribution which presents the bolometric response of the graphene device.
	Together with experimental data, the Fig.~\ref{fig2}b also presents fits to standard bolometric equation $P_{\rm GR}=\sigma(T_{\rm GR}^\beta-T_0^\beta)$~\cite{gantmakher_experimental_1974}, which is based on the power balance condition between the graphene strip and its environment. If the environment is perfectly thermalized,  $T_0$ matches the fridge temperature $T$, but it is not uncommon that below 100 mK the effective environment temperature does not follow $T$, see more in the section ‘Discussion and Outlook’. 
	For all fridge temperatures presented in Fig.~\ref{fig2}b the fit, optimized for intermediate power range $(10^{-15}$–$10^{-13})$~W, returns the same values for parameters $\sigma=2\times 10^{-10}$~${\rm W}/{\rm K}^\beta$ and $\beta=4.7$; at higher temperatures the exponent slowly raises up to $\beta=4.8$ at 200~mK (more arguments in favour of $\beta \approx 5$ are presented is Supplementary sec. S5). We note that at zero power the plots taken at 60~mK and above saturate to the fridge temperature, consistent with graphene being perfectly thermalized to the substrate. However, at the lower temperatures, the graphene temperature saturates to 50~mK regardless of the fridge temperature (for a full data set taken at different fridge temperatures consult the Supplementary sec. S6).

	At high powers/temperatures we observe a deviation from $T^{4.7}$  power dependence, apparently linked to the leakage of heat through quasiparticles in Al superconducting leads~\cite{viisanen_anomalous_2018}. 
	
	\section*{Electron-phonon coupling and Noise Equivalent power}
	Having the numbers for $\sigma,~\beta$, and $T_0$ in  $P_{\rm GR}=\sigma(T_{\rm GR}^\beta-T_0^\beta)$, we can compose a plot for thermal conductance $G_{\rm th}~(T_{\rm GR})=dP_{\rm GR}/dT_{\rm GR}=\beta \sigma T_{\rm GR}^{(\beta-1)}$, presented in Fig.~\ref{fig3}a. The data for two graphene strips with the same geometry, $2\times 8 $ $\mu$m$^2$ have been evaluated in two different cryostats; although the lowest temperature was different in the two setups, we see that the thermal conductivity $G_{\rm th}$ in both experiments follows a universal power law $G_{\rm th}\sim T^{3.7}$ (blue line in Fig.~\ref{fig3}a). The power law $G_{\rm th}\sim T^{3.7}$ is consistent with the encouraging observation $P\sim T^{4.7}$, shown in Fig.~\ref{fig2}b. It has much stronger temperature dependence than previously reported electron-phonon couplings in graphene~\cite{efetov_controlling_2010,betz_hot_2012,borzenets_phonon_2013,lara-avila_towards_2019,karimi_electron-phonon_2021,el_fatimy_epitaxial_2016,el_fatimy_effect_2019}.
	
	This observation correlates with another unexpected result, which is the observation of Arrhenius type conductivity for $T< 200$~mK (Fig.~\ref{fig3}b). Altogether, they suggest that we reach new physics in the heat transfer process from the graphene electronic system to the substrate lattice. We attribute this novel heat transfer regime to the onset of strong localization of charge carriers at low temperatures, caused by quantum interference of the charge carriers in graphene in the presence of a specific type of disorder - such a scenario was mentioned in~\cite{mcardle_electron-phonon_2021,chen_electron-phonon_2012}. 
	
	The unusual electron-phonon decoupling observed in this experiment carries profound implications for the performance of graphene as a bolometer. The figure of merit for a bolometer is its Noise Equivalent Power (NEP)~\cite{low_low-temperature_1961,richards_bolometers_1994}, obtained from the standard formula NEP$=\sqrt{4k_B T_{\rm GR}^2 G_{\rm th}}\sim T_{\rm GR}^{2.85}$. We see that the measured electron-phonon coupling allows for the estimate of NEP of about 40~zW$/\sqrt{\rm{Hz}}$ at $T= 40~$mK, still limited by thermalization issues.  The projected NEP for properly thermalized graphene  at 40~mK would thus surpass the best so far reported number 20~zW$/\sqrt{\rm{Hz}}$~\cite{kokkoniemi_nanobolometer_2019}.

    \begin{figure}
		\centering
		\includegraphics [width=\columnwidth] {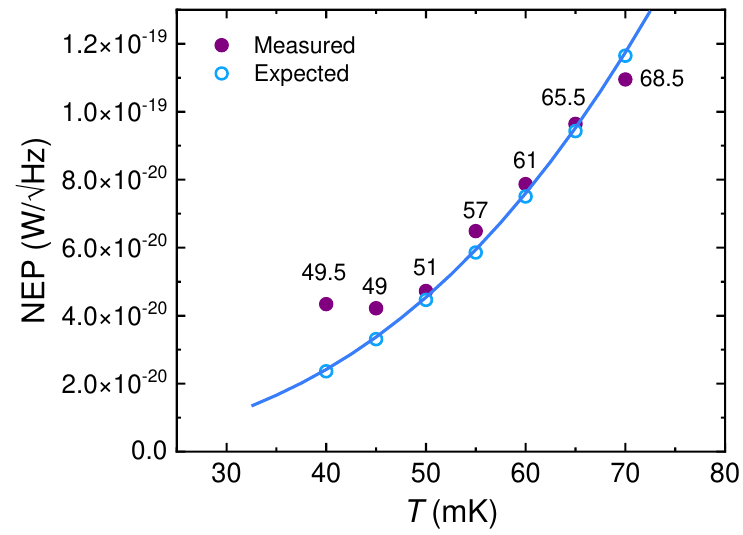}
		\caption{Noise Equivalent Power (NEP) of graphene bolometer as found experimentally (violet circles; the numbers next to markers are for the graphene electronic temperatures $T_0$ - {\it cf }Fig. 2) and as expected for the case of graphene sample thermalized to a fridge temperature (blue circles). The blue line corresponds to $\sqrt{4k_B T^2 G_{\rm th}}$ with parameters $\beta=4.7$ and $\sigma=2\times 10^{-10}$ W/K$^{4.7}$.
			\label{fig4}}
	\end{figure}
	
	\section*{Discussion and outlook}
	
	We have measured electronic temperature of our graphene device with RF transmission technique as a function of applied DC power to graphene and estimated $G_{\rm th}$ from these data. Applying a standard approach for the graphene bolometer performance evaluation in terms of NEP$=\sqrt{4k_BT^2G_{\rm th}}$, we evaluate our graphene bolometer performance as $\sim 40~zW/\sqrt{\rm{Hz}}$ evaluated at 40~mK cryostat base temperature. The observed saturation of graphene temperature at low $T$ cannot be explained by microwave heating or measurement artifacts. Even under conservative estimates, the power dissipated in graphene from the applied microwave excitation is approximately 14~aW, corresponding to a negligible temperature increase of about 4.5~mK, which is insufficient to account for the observed saturation. A detailed analysis of this power estimation is provided in SI section 8. The saturation is therefore more likely due to the heat load from non-thermalized components of the environment such as two-level fluctuators in the substrate dielectric and/or non-equilibrium quasiparticles in aluminum leads. It is a common knowledge that the heat exchange with non-thermalized environment leads to temperature saturation in engineered quantum devices at the level of $50 – 70$~mK~\cite{jin_thermal_2015,kulikov_measuring_2020,sultanov_protocol_2021,lvov_thermometry_2025,paik_observation_2011,burnett_evidence_2014,de_graaf_suppression_2018,lucas_quantum_2023,de_graaf_direct_2017,quintana_observation_2017}, consistent with our observations. Under proper thermalization conditions, the projected NEP for 10~mK is $0.35~z$W$/\sqrt{\rm{Hz}}$; that is not far from the estimate made in our previous work ~\cite{karimi_electron-phonon_2021}. The main difference between the work~\cite{karimi_electron-phonon_2021} and the current measurement is in the lower temperature range, providing an opportunity to convincingly reach the strong localization regime, favoring better bolometric performance at low temperatures. We attribute this result to a synergetic effect of the onset of strong localization of charge carriers in graphene at low temperatures, caused by quantum interference of charge carriers and a consequence of a better, lower noise, less invasive, graphene temperature readout, provided by the RF transmission scheme, implemented in this work. To get a better bolometric performance we need to improve graphene thermalization at low temperatures, which for a bolometric device is especially challenging due to the otherwise beneficial, or even mandatory, thermal decoupling of the graphene bridge from the phonon bath.
    
	In view of prospective calorimetric applications~\cite{satrya_thermal_2025}, it is instructive to evaluate the heat capacity of the graphene bridge. The Fermi energy in graphene is $E_F=\hbar v_F\sqrt{\pi n}$~\cite{castro_neto_electronic_2009, das_sarma_electronic_2011}, {for Fermi velocity $v_F=1\times 10^{6}$ m/s and for the carrier density $n=10^{10}~{\rm cm}^{-2}$ we arrive at $E_F/k_B = 122~K$. Assuming operation temperature $T=50$~mK, we can estimate the number of electrons in a $k_BT$ slice around $E_F$ (the one contributing to the heat capacity) for $S=2 \times 8~\mu{\rm m}^2$ graphene bridge as $2 \frac{k_B T}{E_F} n S\sim 1$. This is quite a remarkable result which shows that as a bolometric sensor, our graphene bridge is equivalent to a quantum dot with a single degree of freedom and the heat capacity $C_e \sim k_B$. The energy resolution is then~\cite{moseley_thermal_1984,lee_graphene-based_2020} $\Delta E =\sqrt{k_B T^2 C_e }\sim k_BT=0.7\times 10^{-24}$~J $=0.7$~yJ, which corresponds to an energy of a single 1 GHz photon.}

	\section*{Conclusions}

	To conclude, we demonstrated that a simple and scalable graphene micrometer-sized strip, cooled to millikelvin temperatures, offers zeptoWatt sensitivity in a bolometric regime, thanks to its temperature-dependent resistance, 
	caused by quantum interference effects, and thermal decoupling. This performance opens the path for scalable high-performance devices in quantum information technology such as click-detector resolving a single 1 GHz photon, 
	or for effective realization of solid-state multi-channel quantum tomography~\cite{pereira_parallel_2023}. Moreover, our technology could be suitable to tackle challenges in cosmology and astronomy, 
	fields where traditional detection of faint radiation has resulted in major breakthroughs~\cite{paolucci_fully_2021}. Finally, the device core graphene bridge, hosting interacting 2D electrons in strongly localized regime,
	is a promising object for studies of thermalization dynamics of non-ergodic many body localized systems~\cite{ueda_quantum_2020}. 

    \section*{Acknowledgements}

    This work at Chalmers was jointly supported by the Chalmers Area of Advance Nano, Chalmers Area of Advanced materials, 2D TECH VINNOVA competence Center (Ref. 2019-00068), Swedish Research Council VR (Contract Nos. 2021-05252), and Knut and Alice Wallenberg foundation via the Wallenberg Center for Quantum Technology (WACQT); A.D. acknowledges support from the Horizon Europe EIC Pathfinder project 101115190 IQARO. Sample fabrication and analysis was supported by Myfab Chalmers and Chalmers Materials Analysis Laboratory (CMAL). This work at Aalto was supported by the Research Council of Finland Centre of Excellence programme grant 336810 and grant 349601 (THEPOW). We sincerely acknowledge the facilities and technical support of Otaniemi Research Infrastructure for Micro and Nanotechnologies (OtaNano) to perform this research. B. K. acknowledges the European Union’s Research and Innovation Programme, Horizon Europe, under the Marie Skłodowska-Curie Grant Agreement No. 101150440 (TcQTD). We are grateful for discussions with Floriana Lombardi, Dmitry S. Golubev, Igor V. Lerner, and  Tord Claeson.
	
\bibliography{graphene_bolometer_refs_v3.bib}

\begin{thebibliography}{49}%
\makeatletter
\providecommand \@ifxundefined [1]{%
 \@ifx{#1\undefined}
}%
\providecommand \@ifnum [1]{%
 \ifnum #1\expandafter \@firstoftwo
 \else \expandafter \@secondoftwo
 \fi
}%
\providecommand \@ifx [1]{%
 \ifx #1\expandafter \@firstoftwo
 \else \expandafter \@secondoftwo
 \fi
}%
\providecommand \natexlab [1]{#1}%
\providecommand \enquote  [1]{``#1''}%
\providecommand \bibnamefont  [1]{#1}%
\providecommand \bibfnamefont [1]{#1}%
\providecommand \citenamefont [1]{#1}%
\providecommand \href@noop [0]{\@secondoftwo}%
\providecommand \href [0]{\begingroup \@sanitize@url \@href}%
\providecommand \@href[1]{\@@startlink{#1}\@@href}%
\providecommand \@@href[1]{\endgroup#1\@@endlink}%
\providecommand \@sanitize@url [0]{\catcode `\\12\catcode `\$12\catcode `\&12\catcode `\#12\catcode `\^12\catcode `\_12\catcode `\%12\relax}%
\providecommand \@@startlink[1]{}%
\providecommand \@@endlink[0]{}%
\providecommand \url  [0]{\begingroup\@sanitize@url \@url }%
\providecommand \@url [1]{\endgroup\@href {#1}{\urlprefix }}%
\providecommand \urlprefix  [0]{URL }%
\providecommand \Eprint [0]{\href }%
\providecommand \doibase [0]{https://doi.org/}%
\providecommand \selectlanguage [0]{\@gobble}%
\providecommand \bibinfo  [0]{\@secondoftwo}%
\providecommand \bibfield  [0]{\@secondoftwo}%
\providecommand \translation [1]{[#1]}%
\providecommand \BibitemOpen [0]{}%
\providecommand \bibitemStop [0]{}%
\providecommand \bibitemNoStop [0]{.\EOS\space}%
\providecommand \EOS [0]{\spacefactor3000\relax}%
\providecommand \BibitemShut  [1]{\csname bibitem#1\endcsname}%
\let\auto@bib@innerbib\@empty
\bibitem [{\citenamefont {Adair}\ \emph {et~al.}(2022)\citenamefont {Adair}, \citenamefont {Altenmüller}, \citenamefont {Anastassopoulos}, \citenamefont {Arguedas~Cuendis}, \citenamefont {Baier}, \citenamefont {Barth}, \citenamefont {Belov}, \citenamefont {Bozicevic}, \citenamefont {Bräuninger}, \citenamefont {Cantatore}, \citenamefont {Caspers}, \citenamefont {Castel}, \citenamefont {\c{C}etin}, \citenamefont {Chung}, \citenamefont {Choi}, \citenamefont {Choi}, \citenamefont {Dafni}, \citenamefont {Davenport}, \citenamefont {Dermenev}, \citenamefont {Desch}, \citenamefont {Döbrich}, \citenamefont {Fischer}, \citenamefont {Funk}, \citenamefont {Galan}, \citenamefont {Gardikiotis}, \citenamefont {Gninenko}, \citenamefont {Golm}, \citenamefont {Hasinoff}, \citenamefont {Hoffmann}, \citenamefont {Díez~Ibáñez}, \citenamefont {Irastorza}, \citenamefont {Jakovčić}, \citenamefont {Kaminski}, \citenamefont {Karuza}, \citenamefont {Krieger}, \citenamefont {Kutlu}, \citenamefont {Lakić}, \citenamefont
  {Laurent}, \citenamefont {Lee}, \citenamefont {Lee}, \citenamefont {Luzón}, \citenamefont {Malbrunot}, \citenamefont {Margalejo}, \citenamefont {Maroudas}, \citenamefont {Miceli}, \citenamefont {Mirallas}, \citenamefont {Obis}, \citenamefont {Özbey}, \citenamefont {Özbozduman}, \citenamefont {Pivovaroff}, \citenamefont {Rosu}, \citenamefont {Ruz}, \citenamefont {Ruiz-Chóliz}, \citenamefont {Schmidt}, \citenamefont {Schumann}, \citenamefont {Semertzidis}, \citenamefont {Solanki}, \citenamefont {Stewart}, \citenamefont {Tsagris}, \citenamefont {Vafeiadis}, \citenamefont {Vogel}, \citenamefont {Vretenar}, \citenamefont {Youn},\ and\ \citenamefont {Zioutas}}]{adair_search_2022}%
  \BibitemOpen
  \bibfield  {author} {\bibinfo {author} {\bibfnamefont {C.~M.}\ \bibnamefont {Adair}}, \bibinfo {author} {\bibfnamefont {K.}~\bibnamefont {Altenmüller}}, \bibinfo {author} {\bibfnamefont {V.}~\bibnamefont {Anastassopoulos}}, \bibinfo {author} {\bibfnamefont {S.}~\bibnamefont {Arguedas~Cuendis}}, \bibinfo {author} {\bibfnamefont {J.}~\bibnamefont {Baier}}, \bibinfo {author} {\bibfnamefont {K.}~\bibnamefont {Barth}}, \bibinfo {author} {\bibfnamefont {A.}~\bibnamefont {Belov}}, \bibinfo {author} {\bibfnamefont {D.}~\bibnamefont {Bozicevic}}, \bibinfo {author} {\bibfnamefont {H.}~\bibnamefont {Bräuninger}}, \bibinfo {author} {\bibfnamefont {G.}~\bibnamefont {Cantatore}}, \bibinfo {author} {\bibfnamefont {F.}~\bibnamefont {Caspers}}, \bibinfo {author} {\bibfnamefont {J.~F.}\ \bibnamefont {Castel}}, \bibinfo {author} {\bibfnamefont {S.~A.}\ \bibnamefont {\c{C}etin}}, \bibinfo {author} {\bibfnamefont {W.}~\bibnamefont {Chung}}, \bibinfo {author} {\bibfnamefont {H.}~\bibnamefont {Choi}}, \bibinfo {author}
  {\bibfnamefont {J.}~\bibnamefont {Choi}}, \bibinfo {author} {\bibfnamefont {T.}~\bibnamefont {Dafni}}, \bibinfo {author} {\bibfnamefont {M.}~\bibnamefont {Davenport}}, \bibinfo {author} {\bibfnamefont {A.}~\bibnamefont {Dermenev}}, \bibinfo {author} {\bibfnamefont {K.}~\bibnamefont {Desch}}, \bibinfo {author} {\bibfnamefont {B.}~\bibnamefont {Döbrich}}, \bibinfo {author} {\bibfnamefont {H.}~\bibnamefont {Fischer}}, \bibinfo {author} {\bibfnamefont {W.}~\bibnamefont {Funk}}, \bibinfo {author} {\bibfnamefont {J.}~\bibnamefont {Galan}}, \bibinfo {author} {\bibfnamefont {A.}~\bibnamefont {Gardikiotis}}, \bibinfo {author} {\bibfnamefont {S.}~\bibnamefont {Gninenko}}, \bibinfo {author} {\bibfnamefont {J.}~\bibnamefont {Golm}}, \bibinfo {author} {\bibfnamefont {M.~D.}\ \bibnamefont {Hasinoff}}, \bibinfo {author} {\bibfnamefont {D.~H.~H.}\ \bibnamefont {Hoffmann}}, \bibinfo {author} {\bibfnamefont {D.}~\bibnamefont {Díez~Ibáñez}}, \bibinfo {author} {\bibfnamefont {I.~G.}\ \bibnamefont {Irastorza}}, \bibinfo
  {author} {\bibfnamefont {K.}~\bibnamefont {Jakovčić}}, \bibinfo {author} {\bibfnamefont {J.}~\bibnamefont {Kaminski}}, \bibinfo {author} {\bibfnamefont {M.}~\bibnamefont {Karuza}}, \bibinfo {author} {\bibfnamefont {C.}~\bibnamefont {Krieger}}, \bibinfo {author} {\bibfnamefont {c.}~\bibnamefont {Kutlu}}, \bibinfo {author} {\bibfnamefont {B.}~\bibnamefont {Lakić}}, \bibinfo {author} {\bibfnamefont {J.~M.}\ \bibnamefont {Laurent}}, \bibinfo {author} {\bibfnamefont {J.}~\bibnamefont {Lee}}, \bibinfo {author} {\bibfnamefont {S.}~\bibnamefont {Lee}}, \bibinfo {author} {\bibfnamefont {G.}~\bibnamefont {Luzón}}, \bibinfo {author} {\bibfnamefont {C.}~\bibnamefont {Malbrunot}}, \bibinfo {author} {\bibfnamefont {C.}~\bibnamefont {Margalejo}}, \bibinfo {author} {\bibfnamefont {M.}~\bibnamefont {Maroudas}}, \bibinfo {author} {\bibfnamefont {L.}~\bibnamefont {Miceli}}, \bibinfo {author} {\bibfnamefont {H.}~\bibnamefont {Mirallas}}, \bibinfo {author} {\bibfnamefont {L.}~\bibnamefont {Obis}}, \bibinfo {author}
  {\bibfnamefont {A.}~\bibnamefont {Özbey}}, \bibinfo {author} {\bibfnamefont {K.}~\bibnamefont {Özbozduman}}, \bibinfo {author} {\bibfnamefont {M.~J.}\ \bibnamefont {Pivovaroff}}, \bibinfo {author} {\bibfnamefont {M.}~\bibnamefont {Rosu}}, \bibinfo {author} {\bibfnamefont {J.}~\bibnamefont {Ruz}}, \bibinfo {author} {\bibfnamefont {E.}~\bibnamefont {Ruiz-Chóliz}}, \bibinfo {author} {\bibfnamefont {S.}~\bibnamefont {Schmidt}}, \bibinfo {author} {\bibfnamefont {M.}~\bibnamefont {Schumann}}, \bibinfo {author} {\bibfnamefont {Y.~K.}\ \bibnamefont {Semertzidis}}, \bibinfo {author} {\bibfnamefont {S.~K.}\ \bibnamefont {Solanki}}, \bibinfo {author} {\bibfnamefont {L.}~\bibnamefont {Stewart}}, \bibinfo {author} {\bibfnamefont {I.}~\bibnamefont {Tsagris}}, \bibinfo {author} {\bibfnamefont {T.}~\bibnamefont {Vafeiadis}}, \bibinfo {author} {\bibfnamefont {J.~K.}\ \bibnamefont {Vogel}}, \bibinfo {author} {\bibfnamefont {M.}~\bibnamefont {Vretenar}}, \bibinfo {author} {\bibfnamefont {S.}~\bibnamefont {Youn}},\ and\
  \bibinfo {author} {\bibfnamefont {K.}~\bibnamefont {Zioutas}},\ }\bibfield  {title} {\bibinfo {title} {Search for {Dark} {Matter} {Axions} with {CAST}-{CAPP}},\ }\href {https://doi.org/10.1038/s41467-022-33913-6} {\bibfield  {journal} {\bibinfo  {journal} {Nat. Commun.}\ }\textbf {\bibinfo {volume} {13}},\ \bibinfo {pages} {6180} (\bibinfo {year} {2022})}\BibitemShut {NoStop}%
\bibitem [{\citenamefont {Braine}\ \emph {et~al.}(2020)\citenamefont {Braine}, \citenamefont {Cervantes}, \citenamefont {Crisosto}, \citenamefont {Du}, \citenamefont {Kimes}, \citenamefont {Rosenberg}, \citenamefont {Rybka}, \citenamefont {Yang}, \citenamefont {Bowring}, \citenamefont {Chou}, \citenamefont {Khatiwada}, \citenamefont {Sonnenschein}, \citenamefont {Wester}, \citenamefont {Carosi}, \citenamefont {Woollett}, \citenamefont {Duffy}, \citenamefont {Bradley}, \citenamefont {Boutan}, \citenamefont {Jones}, \citenamefont {LaRoque}, \citenamefont {Oblath}, \citenamefont {Taubman}, \citenamefont {Clarke}, \citenamefont {Dove}, \citenamefont {Eddins}, \citenamefont {O’Kelley}, \citenamefont {Nawaz}, \citenamefont {Siddiqi}, \citenamefont {Stevenson}, \citenamefont {Agrawal}, \citenamefont {Dixit}, \citenamefont {Gleason}, \citenamefont {Jois}, \citenamefont {Sikivie}, \citenamefont {Solomon}, \citenamefont {Sullivan}, \citenamefont {Tanner}, \citenamefont {Lentz}, \citenamefont {Daw}, \citenamefont
  {Buckley}, \citenamefont {Harrington}, \citenamefont {Henriksen}, \citenamefont {Murch},\ and\ \citenamefont {{ADMX Collaboration}}}]{braine_extended_2020}%
  \BibitemOpen
  \bibfield  {author} {\bibinfo {author} {\bibfnamefont {T.}~\bibnamefont {Braine}}, \bibinfo {author} {\bibfnamefont {R.}~\bibnamefont {Cervantes}}, \bibinfo {author} {\bibfnamefont {N.}~\bibnamefont {Crisosto}}, \bibinfo {author} {\bibfnamefont {N.}~\bibnamefont {Du}}, \bibinfo {author} {\bibfnamefont {S.}~\bibnamefont {Kimes}}, \bibinfo {author} {\bibfnamefont {L.~J.}\ \bibnamefont {Rosenberg}}, \bibinfo {author} {\bibfnamefont {G.}~\bibnamefont {Rybka}}, \bibinfo {author} {\bibfnamefont {J.}~\bibnamefont {Yang}}, \bibinfo {author} {\bibfnamefont {D.}~\bibnamefont {Bowring}}, \bibinfo {author} {\bibfnamefont {A.~S.}\ \bibnamefont {Chou}}, \bibinfo {author} {\bibfnamefont {R.}~\bibnamefont {Khatiwada}}, \bibinfo {author} {\bibfnamefont {A.}~\bibnamefont {Sonnenschein}}, \bibinfo {author} {\bibfnamefont {W.}~\bibnamefont {Wester}}, \bibinfo {author} {\bibfnamefont {G.}~\bibnamefont {Carosi}}, \bibinfo {author} {\bibfnamefont {N.}~\bibnamefont {Woollett}}, \bibinfo {author} {\bibfnamefont {L.~D.}\
  \bibnamefont {Duffy}}, \bibinfo {author} {\bibfnamefont {R.}~\bibnamefont {Bradley}}, \bibinfo {author} {\bibfnamefont {C.}~\bibnamefont {Boutan}}, \bibinfo {author} {\bibfnamefont {M.}~\bibnamefont {Jones}}, \bibinfo {author} {\bibfnamefont {B.~H.}\ \bibnamefont {LaRoque}}, \bibinfo {author} {\bibfnamefont {N.~S.}\ \bibnamefont {Oblath}}, \bibinfo {author} {\bibfnamefont {M.~S.}\ \bibnamefont {Taubman}}, \bibinfo {author} {\bibfnamefont {J.}~\bibnamefont {Clarke}}, \bibinfo {author} {\bibfnamefont {A.}~\bibnamefont {Dove}}, \bibinfo {author} {\bibfnamefont {A.}~\bibnamefont {Eddins}}, \bibinfo {author} {\bibfnamefont {S.~R.}\ \bibnamefont {O’Kelley}}, \bibinfo {author} {\bibfnamefont {S.}~\bibnamefont {Nawaz}}, \bibinfo {author} {\bibfnamefont {I.}~\bibnamefont {Siddiqi}}, \bibinfo {author} {\bibfnamefont {N.}~\bibnamefont {Stevenson}}, \bibinfo {author} {\bibfnamefont {A.}~\bibnamefont {Agrawal}}, \bibinfo {author} {\bibfnamefont {A.~V.}\ \bibnamefont {Dixit}}, \bibinfo {author} {\bibfnamefont {J.~R.}\
  \bibnamefont {Gleason}}, \bibinfo {author} {\bibfnamefont {S.}~\bibnamefont {Jois}}, \bibinfo {author} {\bibfnamefont {P.}~\bibnamefont {Sikivie}}, \bibinfo {author} {\bibfnamefont {J.~A.}\ \bibnamefont {Solomon}}, \bibinfo {author} {\bibfnamefont {N.~S.}\ \bibnamefont {Sullivan}}, \bibinfo {author} {\bibfnamefont {D.~B.}\ \bibnamefont {Tanner}}, \bibinfo {author} {\bibfnamefont {E.}~\bibnamefont {Lentz}}, \bibinfo {author} {\bibfnamefont {E.~J.}\ \bibnamefont {Daw}}, \bibinfo {author} {\bibfnamefont {J.~H.}\ \bibnamefont {Buckley}}, \bibinfo {author} {\bibfnamefont {P.~M.}\ \bibnamefont {Harrington}}, \bibinfo {author} {\bibfnamefont {E.~A.}\ \bibnamefont {Henriksen}}, \bibinfo {author} {\bibfnamefont {K.~W.}\ \bibnamefont {Murch}},\ and\ \bibinfo {author} {\bibnamefont {{ADMX Collaboration}}},\ }\bibfield  {title} {\bibinfo {title} {Extended {Search} for the {Invisible} {Axion} with the {Axion} {Dark} {Matter} {Experiment}},\ }\href {https://doi.org/10.1103/PhysRevLett.124.101303} {\bibfield  {journal}
  {\bibinfo  {journal} {Phys. Rev. Lett.}\ }\textbf {\bibinfo {volume} {124}},\ \bibinfo {pages} {101303} (\bibinfo {year} {2020})}\BibitemShut {NoStop}%
\bibitem [{\citenamefont {Pankratov}\ \emph {et~al.}(2025)\citenamefont {Pankratov}, \citenamefont {Gordeeva}, \citenamefont {Chiginev}, \citenamefont {Revin}, \citenamefont {Blagodatkin}, \citenamefont {Crescini},\ and\ \citenamefont {Kuzmin}}]{pankratov_detection_2025}%
  \BibitemOpen
  \bibfield  {author} {\bibinfo {author} {\bibfnamefont {A.~L.}\ \bibnamefont {Pankratov}}, \bibinfo {author} {\bibfnamefont {A.~V.}\ \bibnamefont {Gordeeva}}, \bibinfo {author} {\bibfnamefont {A.~V.}\ \bibnamefont {Chiginev}}, \bibinfo {author} {\bibfnamefont {L.~S.}\ \bibnamefont {Revin}}, \bibinfo {author} {\bibfnamefont {A.~V.}\ \bibnamefont {Blagodatkin}}, \bibinfo {author} {\bibfnamefont {N.}~\bibnamefont {Crescini}},\ and\ \bibinfo {author} {\bibfnamefont {L.~S.}\ \bibnamefont {Kuzmin}},\ }\bibfield  {title} {\bibinfo {title} {Detection of single-mode thermal microwave photons using an underdamped {Josephson} junction},\ }\href {https://doi.org/10.1038/s41467-025-56040-4} {\bibfield  {journal} {\bibinfo  {journal} {Nat. Commun.}\ }\textbf {\bibinfo {volume} {16}},\ \bibinfo {pages} {3457} (\bibinfo {year} {2025})}\BibitemShut {NoStop}%
\bibitem [{\citenamefont {Irastorza}\ and\ \citenamefont {Redondo}(2018)}]{irastorza_new_2018}%
  \BibitemOpen
  \bibfield  {author} {\bibinfo {author} {\bibfnamefont {I.~G.}\ \bibnamefont {Irastorza}}\ and\ \bibinfo {author} {\bibfnamefont {J.}~\bibnamefont {Redondo}},\ }\bibfield  {title} {\bibinfo {title} {New experimental approaches in the search for axion-like particles},\ }\href {https://doi.org/10.1016/j.ppnp.2018.05.003} {\bibfield  {journal} {\bibinfo  {journal} {Prog. Part. Nucl. Phys.}\ }\textbf {\bibinfo {volume} {102}},\ \bibinfo {pages} {89} (\bibinfo {year} {2018})}\BibitemShut {NoStop}%
\bibitem [{\citenamefont {Sikivie}(1983)}]{sikivie_experimental_1983}%
  \BibitemOpen
  \bibfield  {author} {\bibinfo {author} {\bibfnamefont {P.}~\bibnamefont {Sikivie}},\ }\bibfield  {title} {\bibinfo {title} {Experimental {Tests} of the "{Invisible}" {Axion}},\ }\href {https://doi.org/10.1103/PhysRevLett.51.1415} {\bibfield  {journal} {\bibinfo  {journal} {Phys. Rev. Lett.}\ }\textbf {\bibinfo {volume} {51}},\ \bibinfo {pages} {1415} (\bibinfo {year} {1983})}\BibitemShut {NoStop}%
\bibitem [{\citenamefont {Shokair}\ \emph {et~al.}(2014)\citenamefont {Shokair}, \citenamefont {Root}, \citenamefont {Van~Bibber}, \citenamefont {Brubaker}, \citenamefont {Gurevich}, \citenamefont {Cahn}, \citenamefont {Lamoreaux}, \citenamefont {Anil}, \citenamefont {Lehnert}, \citenamefont {Mitchell}, \citenamefont {Reed},\ and\ \citenamefont {Carosi}}]{shokair_future_2014}%
  \BibitemOpen
  \bibfield  {author} {\bibinfo {author} {\bibfnamefont {T.~M.}\ \bibnamefont {Shokair}}, \bibinfo {author} {\bibfnamefont {J.}~\bibnamefont {Root}}, \bibinfo {author} {\bibfnamefont {K.~A.}\ \bibnamefont {Van~Bibber}}, \bibinfo {author} {\bibfnamefont {B.}~\bibnamefont {Brubaker}}, \bibinfo {author} {\bibfnamefont {Y.~V.}\ \bibnamefont {Gurevich}}, \bibinfo {author} {\bibfnamefont {S.~B.}\ \bibnamefont {Cahn}}, \bibinfo {author} {\bibfnamefont {S.~K.}\ \bibnamefont {Lamoreaux}}, \bibinfo {author} {\bibfnamefont {M.~A.}\ \bibnamefont {Anil}}, \bibinfo {author} {\bibfnamefont {K.~W.}\ \bibnamefont {Lehnert}}, \bibinfo {author} {\bibfnamefont {B.~K.}\ \bibnamefont {Mitchell}}, \bibinfo {author} {\bibfnamefont {A.}~\bibnamefont {Reed}},\ and\ \bibinfo {author} {\bibfnamefont {G.}~\bibnamefont {Carosi}},\ }\bibfield  {title} {\bibinfo {title} {Future directions in the microwave cavity search for dark matter axions},\ }\href {https://doi.org/10.1142/S0217751X14430040} {\bibfield  {journal} {\bibinfo  {journal}
  {Int. J. Mod. Phys. A}\ }\textbf {\bibinfo {volume} {29}},\ \bibinfo {pages} {1443004} (\bibinfo {year} {2014})}\BibitemShut {NoStop}%
\bibitem [{\citenamefont {Karimi}\ and\ \citenamefont {Pekola}(2020)}]{karimi_quantum_2020}%
  \BibitemOpen
  \bibfield  {author} {\bibinfo {author} {\bibfnamefont {B.}~\bibnamefont {Karimi}}\ and\ \bibinfo {author} {\bibfnamefont {J.~P.}\ \bibnamefont {Pekola}},\ }\bibfield  {title} {\bibinfo {title} {Quantum {Trajectory} {Analysis} of {Single} {Microwave} {Photon} {Detection} by {Nanocalorimetry}},\ }\href@noop {} {\bibfield  {journal} {\bibinfo  {journal} {Phys. Rev. Lett.}\ }\textbf {\bibinfo {volume} {124}},\ \bibinfo {pages} {170601} (\bibinfo {year} {2020})}\BibitemShut {NoStop}%
\bibitem [{\citenamefont {Gunyhó}\ \emph {et~al.}(2024)\citenamefont {Gunyhó}, \citenamefont {Kundu}, \citenamefont {Ma}, \citenamefont {Liu}, \citenamefont {Niemelä}, \citenamefont {Catto}, \citenamefont {Vadimov}, \citenamefont {Vesterinen}, \citenamefont {Singh}, \citenamefont {Chen},\ and\ \citenamefont {Möttönen}}]{gunyho_single-shot_2024}%
  \BibitemOpen
  \bibfield  {author} {\bibinfo {author} {\bibfnamefont {A.~M.}\ \bibnamefont {Gunyhó}}, \bibinfo {author} {\bibfnamefont {S.}~\bibnamefont {Kundu}}, \bibinfo {author} {\bibfnamefont {J.}~\bibnamefont {Ma}}, \bibinfo {author} {\bibfnamefont {W.}~\bibnamefont {Liu}}, \bibinfo {author} {\bibfnamefont {S.}~\bibnamefont {Niemelä}}, \bibinfo {author} {\bibfnamefont {G.}~\bibnamefont {Catto}}, \bibinfo {author} {\bibfnamefont {V.}~\bibnamefont {Vadimov}}, \bibinfo {author} {\bibfnamefont {V.}~\bibnamefont {Vesterinen}}, \bibinfo {author} {\bibfnamefont {P.}~\bibnamefont {Singh}}, \bibinfo {author} {\bibfnamefont {Q.}~\bibnamefont {Chen}},\ and\ \bibinfo {author} {\bibfnamefont {M.}~\bibnamefont {Möttönen}},\ }\bibfield  {title} {\bibinfo {title} {Single-shot readout of a superconducting qubit using a thermal detector},\ }\href@noop {} {\bibfield  {journal} {\bibinfo  {journal} {Nat. Electron.}\ }\textbf {\bibinfo {volume} {7}},\ \bibinfo {pages} {288} (\bibinfo {year} {2024})}\BibitemShut {NoStop}%
\bibitem [{\citenamefont {McKitterick}\ \emph {et~al.}(2015)\citenamefont {McKitterick}, \citenamefont {Prober}, \citenamefont {Vora},\ and\ \citenamefont {Du}}]{mckitterick_ultrasensitive_2015}%
  \BibitemOpen
  \bibfield  {author} {\bibinfo {author} {\bibfnamefont {C.~B.}\ \bibnamefont {McKitterick}}, \bibinfo {author} {\bibfnamefont {D.~E.}\ \bibnamefont {Prober}}, \bibinfo {author} {\bibfnamefont {H.}~\bibnamefont {Vora}},\ and\ \bibinfo {author} {\bibfnamefont {X.}~\bibnamefont {Du}},\ }\bibfield  {title} {\bibinfo {title} {Ultrasensitive graphene far-infrared power detectors},\ }\href {https://doi.org/10.1088/0953-8984/27/16/164203} {\bibfield  {journal} {\bibinfo  {journal} {J. Phys.: Condens. Matter}\ }\textbf {\bibinfo {volume} {27}},\ \bibinfo {pages} {164203} (\bibinfo {year} {2015})}\BibitemShut {NoStop}%
\bibitem [{\citenamefont {Efetov}\ and\ \citenamefont {Kim}(2010)}]{efetov_controlling_2010}%
  \BibitemOpen
  \bibfield  {author} {\bibinfo {author} {\bibfnamefont {D.~K.}\ \bibnamefont {Efetov}}\ and\ \bibinfo {author} {\bibfnamefont {P.}~\bibnamefont {Kim}},\ }\bibfield  {title} {\bibinfo {title} {Controlling {Electron}-{Phonon} {Interactions} in {Graphene} at {Ultrahigh} {Carrier} {Densities}},\ }\href {https://doi.org/10.1103/PhysRevLett.105.256805} {\bibfield  {journal} {\bibinfo  {journal} {Phys. Rev. Lett.}\ }\textbf {\bibinfo {volume} {105}},\ \bibinfo {pages} {256805} (\bibinfo {year} {2010})}\BibitemShut {NoStop}%
\bibitem [{\citenamefont {Kokkoniemi}\ \emph {et~al.}(2020)\citenamefont {Kokkoniemi}, \citenamefont {Girard}, \citenamefont {Hazra}, \citenamefont {Laitinen}, \citenamefont {Govenius}, \citenamefont {Lake}, \citenamefont {Sallinen}, \citenamefont {Vesterinen}, \citenamefont {Partanen}, \citenamefont {Tan}, \citenamefont {Chan}, \citenamefont {Tan}, \citenamefont {Hakonen},\ and\ \citenamefont {Möttönen}}]{kokkoniemi_bolometer_2020}%
  \BibitemOpen
  \bibfield  {author} {\bibinfo {author} {\bibfnamefont {R.}~\bibnamefont {Kokkoniemi}}, \bibinfo {author} {\bibfnamefont {J.-P.}\ \bibnamefont {Girard}}, \bibinfo {author} {\bibfnamefont {D.}~\bibnamefont {Hazra}}, \bibinfo {author} {\bibfnamefont {A.}~\bibnamefont {Laitinen}}, \bibinfo {author} {\bibfnamefont {J.}~\bibnamefont {Govenius}}, \bibinfo {author} {\bibfnamefont {R.~E.}\ \bibnamefont {Lake}}, \bibinfo {author} {\bibfnamefont {I.}~\bibnamefont {Sallinen}}, \bibinfo {author} {\bibfnamefont {V.}~\bibnamefont {Vesterinen}}, \bibinfo {author} {\bibfnamefont {M.}~\bibnamefont {Partanen}}, \bibinfo {author} {\bibfnamefont {J.~Y.}\ \bibnamefont {Tan}}, \bibinfo {author} {\bibfnamefont {K.~W.}\ \bibnamefont {Chan}}, \bibinfo {author} {\bibfnamefont {K.~Y.}\ \bibnamefont {Tan}}, \bibinfo {author} {\bibfnamefont {P.}~\bibnamefont {Hakonen}},\ and\ \bibinfo {author} {\bibfnamefont {M.}~\bibnamefont {Möttönen}},\ }\bibfield  {title} {\bibinfo {title} {Bolometer operating at the threshold for circuit quantum
  electrodynamics},\ }\href {https://doi.org/https://doi.org/10.1038/s41586-020-2753-3} {\bibfield  {journal} {\bibinfo  {journal} {Nature}\ }\textbf {\bibinfo {volume} {586}},\ \bibinfo {pages} {47} (\bibinfo {year} {2020})}\BibitemShut {NoStop}%
\bibitem [{\citenamefont {He}\ \emph {et~al.}(2022)\citenamefont {He}, \citenamefont {Cedergren}, \citenamefont {Shetty}, \citenamefont {Lara-Avila}, \citenamefont {Kubatkin}, \citenamefont {Bergsten},\ and\ \citenamefont {Eklund}}]{he_accurate_2022}%
  \BibitemOpen
  \bibfield  {author} {\bibinfo {author} {\bibfnamefont {H.}~\bibnamefont {He}}, \bibinfo {author} {\bibfnamefont {K.}~\bibnamefont {Cedergren}}, \bibinfo {author} {\bibfnamefont {N.}~\bibnamefont {Shetty}}, \bibinfo {author} {\bibfnamefont {S.}~\bibnamefont {Lara-Avila}}, \bibinfo {author} {\bibfnamefont {S.}~\bibnamefont {Kubatkin}}, \bibinfo {author} {\bibfnamefont {T.}~\bibnamefont {Bergsten}},\ and\ \bibinfo {author} {\bibfnamefont {G.}~\bibnamefont {Eklund}},\ }\bibfield  {title} {\bibinfo {title} {Accurate graphene quantum {Hall} arrays for the new {International} {System} of {Units}},\ }\href {https://doi.org/https://doi.org/10.1038/s41467-022-34680-0} {\bibfield  {journal} {\bibinfo  {journal} {Nat. Commun.}\ }\textbf {\bibinfo {volume} {13}},\ \bibinfo {pages} {6933} (\bibinfo {year} {2022})}\BibitemShut {NoStop}%
\bibitem [{\citenamefont {McCann}\ \emph {et~al.}(2006)\citenamefont {McCann}, \citenamefont {Kechedzhi}, \citenamefont {Fal’ko}, \citenamefont {Suzuura}, \citenamefont {Ando},\ and\ \citenamefont {Altshuler}}]{mccann_weak-localization_2006}%
  \BibitemOpen
  \bibfield  {author} {\bibinfo {author} {\bibfnamefont {E.}~\bibnamefont {McCann}}, \bibinfo {author} {\bibfnamefont {K.}~\bibnamefont {Kechedzhi}}, \bibinfo {author} {\bibfnamefont {V.~I.}\ \bibnamefont {Fal’ko}}, \bibinfo {author} {\bibfnamefont {H.}~\bibnamefont {Suzuura}}, \bibinfo {author} {\bibfnamefont {T.}~\bibnamefont {Ando}},\ and\ \bibinfo {author} {\bibfnamefont {B.~L.}\ \bibnamefont {Altshuler}},\ }\bibfield  {title} {\bibinfo {title} {Weak-{Localization} {Magnetoresistance} and {Valley} {Symmetry} in {Graphene}},\ }\href {https://doi.org/10.1103/PhysRevLett.97.146805} {\bibfield  {journal} {\bibinfo  {journal} {Phys. Rev. Lett.}\ }\textbf {\bibinfo {volume} {97}},\ \bibinfo {pages} {146805} (\bibinfo {year} {2006})}\BibitemShut {NoStop}%
\bibitem [{\citenamefont {Lara-Avila}\ \emph {et~al.}(2011)\citenamefont {Lara-Avila}, \citenamefont {Tzalenchuk}, \citenamefont {Kubatkin}, \citenamefont {Yakimova}, \citenamefont {Janssen}, \citenamefont {Cedergren}, \citenamefont {Bergsten},\ and\ \citenamefont {Fal'ko}}]{lara-avila_disordered_2011}%
  \BibitemOpen
  \bibfield  {author} {\bibinfo {author} {\bibfnamefont {S.}~\bibnamefont {Lara-Avila}}, \bibinfo {author} {\bibfnamefont {A.}~\bibnamefont {Tzalenchuk}}, \bibinfo {author} {\bibfnamefont {S.}~\bibnamefont {Kubatkin}}, \bibinfo {author} {\bibfnamefont {R.}~\bibnamefont {Yakimova}}, \bibinfo {author} {\bibfnamefont {T.~J. B.~M.}\ \bibnamefont {Janssen}}, \bibinfo {author} {\bibfnamefont {K.}~\bibnamefont {Cedergren}}, \bibinfo {author} {\bibfnamefont {T.}~\bibnamefont {Bergsten}},\ and\ \bibinfo {author} {\bibfnamefont {V.}~\bibnamefont {Fal'ko}},\ }\bibfield  {title} {\bibinfo {title} {Disordered {Fermi} {Liquid} in {Epitaxial} {Graphene} from {Quantum} {Transport} {Measurements}},\ }\href {https://doi.org/10.1103/PhysRevLett.107.166602} {\bibfield  {journal} {\bibinfo  {journal} {Phys. Rev. Lett.}\ }\textbf {\bibinfo {volume} {107}},\ \bibinfo {pages} {166602} (\bibinfo {year} {2011})}\BibitemShut {NoStop}%
\bibitem [{\citenamefont {Ponomarenko}\ \emph {et~al.}(2011)\citenamefont {Ponomarenko}, \citenamefont {Geim}, \citenamefont {Zhukov}, \citenamefont {Jalil}, \citenamefont {Morozov}, \citenamefont {Novoselov}, \citenamefont {Grigorieva}, \citenamefont {Hill}, \citenamefont {Cheianov}, \citenamefont {Faĺko}, \citenamefont {Watanabe}, \citenamefont {Taniguchi},\ and\ \citenamefont {Gorbachev}}]{ponomarenko_tunable_2011}%
  \BibitemOpen
  \bibfield  {author} {\bibinfo {author} {\bibfnamefont {L.~A.}\ \bibnamefont {Ponomarenko}}, \bibinfo {author} {\bibfnamefont {A.~K.}\ \bibnamefont {Geim}}, \bibinfo {author} {\bibfnamefont {A.~A.}\ \bibnamefont {Zhukov}}, \bibinfo {author} {\bibfnamefont {R.}~\bibnamefont {Jalil}}, \bibinfo {author} {\bibfnamefont {S.~V.}\ \bibnamefont {Morozov}}, \bibinfo {author} {\bibfnamefont {K.~S.}\ \bibnamefont {Novoselov}}, \bibinfo {author} {\bibfnamefont {I.~V.}\ \bibnamefont {Grigorieva}}, \bibinfo {author} {\bibfnamefont {E.~H.}\ \bibnamefont {Hill}}, \bibinfo {author} {\bibfnamefont {V.~V.}\ \bibnamefont {Cheianov}}, \bibinfo {author} {\bibfnamefont {V.~I.}\ \bibnamefont {Faĺko}}, \bibinfo {author} {\bibfnamefont {K.}~\bibnamefont {Watanabe}}, \bibinfo {author} {\bibfnamefont {T.}~\bibnamefont {Taniguchi}},\ and\ \bibinfo {author} {\bibfnamefont {a.~R.~V.}\ \bibnamefont {Gorbachev}},\ }\bibfield  {title} {\bibinfo {title} {Tunable metal–insulator transition in double-layer graphene heterostructures},\ }\href
  {https://doi.org/https://doi.org/10.1038/nphys2114} {\bibfield  {journal} {\bibinfo  {journal} {Nat. Phys.}\ }\textbf {\bibinfo {volume} {7}},\ \bibinfo {pages} {958} (\bibinfo {year} {2011})}\BibitemShut {NoStop}%
\bibitem [{\citenamefont {Roukes}(1999)}]{roukes_yoctocalorimetry_1999}%
  \BibitemOpen
  \bibfield  {author} {\bibinfo {author} {\bibfnamefont {M.~L.}\ \bibnamefont {Roukes}},\ }\bibfield  {title} {\bibinfo {title} {Yoctocalorimetry: phonon counting in nanostructures},\ }\href {https://doi.org/10.1016/S0921-4526(98)01482-3} {\bibfield  {journal} {\bibinfo  {journal} {Physica B: Condensed Matter}\ }\textbf {\bibinfo {volume} {263-264}},\ \bibinfo {pages} {1} (\bibinfo {year} {1999})}\BibitemShut {NoStop}%
\bibitem [{\citenamefont {He}\ \emph {et~al.}(2018)\citenamefont {He}, \citenamefont {Kim}, \citenamefont {Danilov}, \citenamefont {Montemurro}, \citenamefont {Yu}, \citenamefont {Park}, \citenamefont {Lombardi}, \citenamefont {Bauch}, \citenamefont {Moth-Poulsen}, \citenamefont {Iakimov}, \citenamefont {Yakimova}, \citenamefont {Malmberg}, \citenamefont {Müller}, \citenamefont {Kubatkin},\ and\ \citenamefont {Lara-Avila}}]{he_uniform_2018}%
  \BibitemOpen
  \bibfield  {author} {\bibinfo {author} {\bibfnamefont {H.}~\bibnamefont {He}}, \bibinfo {author} {\bibfnamefont {K.~H.}\ \bibnamefont {Kim}}, \bibinfo {author} {\bibfnamefont {A.}~\bibnamefont {Danilov}}, \bibinfo {author} {\bibfnamefont {D.}~\bibnamefont {Montemurro}}, \bibinfo {author} {\bibfnamefont {L.}~\bibnamefont {Yu}}, \bibinfo {author} {\bibfnamefont {Y.~W.}\ \bibnamefont {Park}}, \bibinfo {author} {\bibfnamefont {F.}~\bibnamefont {Lombardi}}, \bibinfo {author} {\bibfnamefont {T.}~\bibnamefont {Bauch}}, \bibinfo {author} {\bibfnamefont {K.}~\bibnamefont {Moth-Poulsen}}, \bibinfo {author} {\bibfnamefont {T.}~\bibnamefont {Iakimov}}, \bibinfo {author} {\bibfnamefont {R.}~\bibnamefont {Yakimova}}, \bibinfo {author} {\bibfnamefont {P.}~\bibnamefont {Malmberg}}, \bibinfo {author} {\bibfnamefont {C.}~\bibnamefont {Müller}}, \bibinfo {author} {\bibfnamefont {S.}~\bibnamefont {Kubatkin}},\ and\ \bibinfo {author} {\bibfnamefont {S.}~\bibnamefont {Lara-Avila}},\ }\bibfield  {title} {\bibinfo {title}
  {Uniform doping of graphene close to the {Dirac} point by polymer-assisted assembly of molecular dopants},\ }\href {https://doi.org/10.1038/s41467-018-06352-5} {\bibfield  {journal} {\bibinfo  {journal} {Nat. Commun.}\ }\textbf {\bibinfo {volume} {9}},\ \bibinfo {pages} {3956} (\bibinfo {year} {2018})}\BibitemShut {NoStop}%
\bibitem [{\citenamefont {Schoelkopf}\ \emph {et~al.}(1998)\citenamefont {Schoelkopf}, \citenamefont {Wahlgren}, \citenamefont {Kozhevnikov}, \citenamefont {Delsing},\ and\ \citenamefont {Prober}}]{schoelkopf_radio-frequency_1998}%
  \BibitemOpen
  \bibfield  {author} {\bibinfo {author} {\bibfnamefont {R.~J.}\ \bibnamefont {Schoelkopf}}, \bibinfo {author} {\bibfnamefont {P.}~\bibnamefont {Wahlgren}}, \bibinfo {author} {\bibfnamefont {A.~A.}\ \bibnamefont {Kozhevnikov}}, \bibinfo {author} {\bibfnamefont {P.}~\bibnamefont {Delsing}},\ and\ \bibinfo {author} {\bibfnamefont {D.~E.}\ \bibnamefont {Prober}},\ }\bibfield  {title} {\bibinfo {title} {The {Radio}-{Frequency} {Single}-{Electron} {Transistor} ({RF}-{SET}): {A} {Fast} and {Ultrasensitive} {Electrometer}},\ }\href {https://doi.org/https://www.science.org/doi/10.1126/science.280.5367.1238} {\bibfield  {journal} {\bibinfo  {journal} {Science}\ }\textbf {\bibinfo {volume} {280}},\ \bibinfo {pages} {1238} (\bibinfo {year} {1998})}\BibitemShut {NoStop}%
\bibitem [{\citenamefont {Gantmakher}(1974)}]{gantmakher_experimental_1974}%
  \BibitemOpen
  \bibfield  {author} {\bibinfo {author} {\bibfnamefont {V.~F.}\ \bibnamefont {Gantmakher}},\ }\bibfield  {title} {\bibinfo {title} {The experimental study of electron-phonon scattering in metals},\ }\href {https://doi.org/10.1088/0034-4885/37/3/001} {\bibfield  {journal} {\bibinfo  {journal} {Rep. Prog. Phys.}\ }\textbf {\bibinfo {volume} {37}},\ \bibinfo {pages} {317} (\bibinfo {year} {1974})}\BibitemShut {NoStop}%
\bibitem [{\citenamefont {Viisanen}\ and\ \citenamefont {Pekola}(2018)}]{viisanen_anomalous_2018}%
  \BibitemOpen
  \bibfield  {author} {\bibinfo {author} {\bibfnamefont {K.~L.}\ \bibnamefont {Viisanen}}\ and\ \bibinfo {author} {\bibfnamefont {J.~P.}\ \bibnamefont {Pekola}},\ }\bibfield  {title} {\bibinfo {title} {Anomalous electronic heat capacity of copper nanowires at sub-{Kelvin} temperatures},\ }\href {https://doi.org/10.1103/PhysRevB.97.115422} {\bibfield  {journal} {\bibinfo  {journal} {Phys. Rev. B}\ }\textbf {\bibinfo {volume} {97}},\ \bibinfo {pages} {115422} (\bibinfo {year} {2018})}\BibitemShut {NoStop}%
\bibitem [{\citenamefont {Betz}\ \emph {et~al.}(2012)\citenamefont {Betz}, \citenamefont {Vialla}, \citenamefont {Brunel}, \citenamefont {Voisin}, \citenamefont {Picher}, \citenamefont {Cavanna}, \citenamefont {Madouri}, \citenamefont {Fève}, \citenamefont {Berroir}, \citenamefont {Plaçais},\ and\ \citenamefont {Pallecchi}}]{betz_hot_2012}%
  \BibitemOpen
  \bibfield  {author} {\bibinfo {author} {\bibfnamefont {A.~C.}\ \bibnamefont {Betz}}, \bibinfo {author} {\bibfnamefont {F.}~\bibnamefont {Vialla}}, \bibinfo {author} {\bibfnamefont {D.}~\bibnamefont {Brunel}}, \bibinfo {author} {\bibfnamefont {C.}~\bibnamefont {Voisin}}, \bibinfo {author} {\bibfnamefont {M.}~\bibnamefont {Picher}}, \bibinfo {author} {\bibfnamefont {A.}~\bibnamefont {Cavanna}}, \bibinfo {author} {\bibfnamefont {A.}~\bibnamefont {Madouri}}, \bibinfo {author} {\bibfnamefont {G.}~\bibnamefont {Fève}}, \bibinfo {author} {\bibfnamefont {J.-M.}\ \bibnamefont {Berroir}}, \bibinfo {author} {\bibfnamefont {B.}~\bibnamefont {Plaçais}},\ and\ \bibinfo {author} {\bibfnamefont {E.}~\bibnamefont {Pallecchi}},\ }\bibfield  {title} {\bibinfo {title} {Hot {Electron} {Cooling} by {Acoustic} {Phonons} in {Graphene}},\ }\href {https://doi.org/10.1103/PhysRevLett.109.056805} {\bibfield  {journal} {\bibinfo  {journal} {Phys. Rev. Lett.}\ }\textbf {\bibinfo {volume} {109}},\ \bibinfo {pages} {056805} (\bibinfo
  {year} {2012})}\BibitemShut {NoStop}%
\bibitem [{\citenamefont {Borzenets}\ \emph {et~al.}(2013)\citenamefont {Borzenets}, \citenamefont {Coskun}, \citenamefont {Mebrahtu}, \citenamefont {Bomze}, \citenamefont {Smirnov},\ and\ \citenamefont {Finkelstein}}]{borzenets_phonon_2013}%
  \BibitemOpen
  \bibfield  {author} {\bibinfo {author} {\bibfnamefont {I.~V.}\ \bibnamefont {Borzenets}}, \bibinfo {author} {\bibfnamefont {U.~C.}\ \bibnamefont {Coskun}}, \bibinfo {author} {\bibfnamefont {H.~T.}\ \bibnamefont {Mebrahtu}}, \bibinfo {author} {\bibfnamefont {Y.~V.}\ \bibnamefont {Bomze}}, \bibinfo {author} {\bibfnamefont {A.~I.}\ \bibnamefont {Smirnov}},\ and\ \bibinfo {author} {\bibfnamefont {G.}~\bibnamefont {Finkelstein}},\ }\bibfield  {title} {\bibinfo {title} {Phonon {Bottleneck} in {Graphene}-{Based} {Josephson} {Junctions} at {Millikelvin} {Temperatures}},\ }\href {https://doi.org/10.1103/PhysRevLett.111.027001} {\bibfield  {journal} {\bibinfo  {journal} {Phys. Rev. Lett.}\ }\textbf {\bibinfo {volume} {111}},\ \bibinfo {pages} {027001} (\bibinfo {year} {2013})}\BibitemShut {NoStop}%
\bibitem [{\citenamefont {Lara-Avila}\ \emph {et~al.}(2019)\citenamefont {Lara-Avila}, \citenamefont {Danilov}, \citenamefont {Golubev}, \citenamefont {He}, \citenamefont {Kim}, \citenamefont {Yakimova}, \citenamefont {Lombardi}, \citenamefont {Bauch}, \citenamefont {Cherednichenko},\ and\ \citenamefont {Kubatkin}}]{lara-avila_towards_2019}%
  \BibitemOpen
  \bibfield  {author} {\bibinfo {author} {\bibfnamefont {S.}~\bibnamefont {Lara-Avila}}, \bibinfo {author} {\bibfnamefont {A.}~\bibnamefont {Danilov}}, \bibinfo {author} {\bibfnamefont {D.}~\bibnamefont {Golubev}}, \bibinfo {author} {\bibfnamefont {H.}~\bibnamefont {He}}, \bibinfo {author} {\bibfnamefont {K.~H.}\ \bibnamefont {Kim}}, \bibinfo {author} {\bibfnamefont {R.}~\bibnamefont {Yakimova}}, \bibinfo {author} {\bibfnamefont {F.}~\bibnamefont {Lombardi}}, \bibinfo {author} {\bibfnamefont {T.}~\bibnamefont {Bauch}}, \bibinfo {author} {\bibfnamefont {S.}~\bibnamefont {Cherednichenko}},\ and\ \bibinfo {author} {\bibfnamefont {S.}~\bibnamefont {Kubatkin}},\ }\bibfield  {title} {\bibinfo {title} {Towards quantum-limited coherent detection of terahertz waves in charge-neutral graphene},\ }\href {https://doi.org/10.1038/s41550-019-0843-7} {\bibfield  {journal} {\bibinfo  {journal} {Nat. Astron.}\ }\textbf {\bibinfo {volume} {3}},\ \bibinfo {pages} {983} (\bibinfo {year} {2019})}\BibitemShut {NoStop}%
\bibitem [{\citenamefont {Karimi}\ \emph {et~al.}(2021)\citenamefont {Karimi}, \citenamefont {He}, \citenamefont {Chang}, \citenamefont {Wang}, \citenamefont {Pekola}, \citenamefont {Yakimova}, \citenamefont {Shetty}, \citenamefont {Peltonen}, \citenamefont {Lara-Avila},\ and\ \citenamefont {Kubatkin}}]{karimi_electron-phonon_2021}%
  \BibitemOpen
  \bibfield  {author} {\bibinfo {author} {\bibfnamefont {B.}~\bibnamefont {Karimi}}, \bibinfo {author} {\bibfnamefont {H.}~\bibnamefont {He}}, \bibinfo {author} {\bibfnamefont {Y.~C.}\ \bibnamefont {Chang}}, \bibinfo {author} {\bibfnamefont {L.}~\bibnamefont {Wang}}, \bibinfo {author} {\bibfnamefont {J.~P.}\ \bibnamefont {Pekola}}, \bibinfo {author} {\bibfnamefont {R.}~\bibnamefont {Yakimova}}, \bibinfo {author} {\bibfnamefont {N.}~\bibnamefont {Shetty}}, \bibinfo {author} {\bibfnamefont {J.~T.}\ \bibnamefont {Peltonen}}, \bibinfo {author} {\bibfnamefont {S.}~\bibnamefont {Lara-Avila}},\ and\ \bibinfo {author} {\bibfnamefont {S.}~\bibnamefont {Kubatkin}},\ }\bibfield  {title} {\bibinfo {title} {Electron-phonon coupling of epigraphene at millikelvin temperatures measured by quantum transport thermometry},\ }\href {https://pubs.aip.org/aip/apl/article/118/10/103102/1061405/Electron-phonon-coupling-of-epigraphene-at} {\bibfield  {journal} {\bibinfo  {journal} {Appl. Phys. Lett.}\ }\textbf {\bibinfo {volume}
  {118}},\ \bibinfo {pages} {103102} (\bibinfo {year} {2021})}\BibitemShut {NoStop}%
\bibitem [{\citenamefont {El~Fatimy}\ \emph {et~al.}(2016)\citenamefont {El~Fatimy}, \citenamefont {Myers-Ward}, \citenamefont {Boyd}, \citenamefont {Daniels}, \citenamefont {Gaskill},\ and\ \citenamefont {Barbara}}]{el_fatimy_epitaxial_2016}%
  \BibitemOpen
  \bibfield  {author} {\bibinfo {author} {\bibfnamefont {A.}~\bibnamefont {El~Fatimy}}, \bibinfo {author} {\bibfnamefont {R.~L.}\ \bibnamefont {Myers-Ward}}, \bibinfo {author} {\bibfnamefont {A.~K.}\ \bibnamefont {Boyd}}, \bibinfo {author} {\bibfnamefont {K.~M.}\ \bibnamefont {Daniels}}, \bibinfo {author} {\bibfnamefont {D.~K.}\ \bibnamefont {Gaskill}},\ and\ \bibinfo {author} {\bibfnamefont {P.}~\bibnamefont {Barbara}},\ }\bibfield  {title} {\bibinfo {title} {Epitaxial graphene quantum dots for high-performance terahertz bolometers},\ }\href {https://doi.org/10.1038/nnano.2015.303} {\bibfield  {journal} {\bibinfo  {journal} {Nat. Nanotechnol.}\ }\textbf {\bibinfo {volume} {11}},\ \bibinfo {pages} {335} (\bibinfo {year} {2016})}\BibitemShut {NoStop}%
\bibitem [{\citenamefont {El~Fatimy}\ \emph {et~al.}(2019)\citenamefont {El~Fatimy}, \citenamefont {Han}, \citenamefont {Quirk}, \citenamefont {St.~Marie}, \citenamefont {Dejarld}, \citenamefont {Myers-Ward}, \citenamefont {Daniels}, \citenamefont {Pavunny}, \citenamefont {Gaskill}, \citenamefont {Aytac}, \citenamefont {Murphy},\ and\ \citenamefont {Barbara}}]{el_fatimy_effect_2019}%
  \BibitemOpen
  \bibfield  {author} {\bibinfo {author} {\bibfnamefont {A.}~\bibnamefont {El~Fatimy}}, \bibinfo {author} {\bibfnamefont {P.}~\bibnamefont {Han}}, \bibinfo {author} {\bibfnamefont {N.}~\bibnamefont {Quirk}}, \bibinfo {author} {\bibfnamefont {L.}~\bibnamefont {St.~Marie}}, \bibinfo {author} {\bibfnamefont {M.~T.}\ \bibnamefont {Dejarld}}, \bibinfo {author} {\bibfnamefont {R.~L.}\ \bibnamefont {Myers-Ward}}, \bibinfo {author} {\bibfnamefont {K.}~\bibnamefont {Daniels}}, \bibinfo {author} {\bibfnamefont {S.}~\bibnamefont {Pavunny}}, \bibinfo {author} {\bibfnamefont {D.~K.}\ \bibnamefont {Gaskill}}, \bibinfo {author} {\bibfnamefont {Y.}~\bibnamefont {Aytac}}, \bibinfo {author} {\bibfnamefont {T.~E.}\ \bibnamefont {Murphy}},\ and\ \bibinfo {author} {\bibfnamefont {P.}~\bibnamefont {Barbara}},\ }\bibfield  {title} {\bibinfo {title} {Effect of defect-induced cooling on graphene hot-electron bolometers},\ }\href {https://doi.org/10.1016/j.carbon.2019.08.019} {\bibfield  {journal} {\bibinfo  {journal} {Carbon}\
  }\textbf {\bibinfo {volume} {154}},\ \bibinfo {pages} {497} (\bibinfo {year} {2019})}\BibitemShut {NoStop}%
\bibitem [{\citenamefont {McArdle}\ and\ \citenamefont {Lerner}(2021)}]{mcardle_electron-phonon_2021}%
  \BibitemOpen
  \bibfield  {author} {\bibinfo {author} {\bibfnamefont {G.}~\bibnamefont {McArdle}}\ and\ \bibinfo {author} {\bibfnamefont {I.~V.}\ \bibnamefont {Lerner}},\ }\bibfield  {title} {\bibinfo {title} {Electron-phonon decoupling in two dimensions},\ }\href {https://doi.org/https://doi.org/10.1038/s41598-021-03668-z} {\bibfield  {journal} {\bibinfo  {journal} {Sci. Rep.}\ }\textbf {\bibinfo {volume} {11}},\ \bibinfo {pages} {24293} (\bibinfo {year} {2021})}\BibitemShut {NoStop}%
\bibitem [{\citenamefont {Chen}\ and\ \citenamefont {Clerk}(2012)}]{chen_electron-phonon_2012}%
  \BibitemOpen
  \bibfield  {author} {\bibinfo {author} {\bibfnamefont {W.}~\bibnamefont {Chen}}\ and\ \bibinfo {author} {\bibfnamefont {A.~A.}\ \bibnamefont {Clerk}},\ }\bibfield  {title} {\bibinfo {title} {Electron-phonon mediated heat flow in disordered graphene},\ }\href {https://doi.org/10.1103/PhysRevB.86.125443} {\bibfield  {journal} {\bibinfo  {journal} {Phys. Rev. B}\ }\textbf {\bibinfo {volume} {86}},\ \bibinfo {pages} {125443} (\bibinfo {year} {2012})}\BibitemShut {NoStop}%
\bibitem [{\citenamefont {Low}(1961)}]{low_low-temperature_1961}%
  \BibitemOpen
  \bibfield  {author} {\bibinfo {author} {\bibfnamefont {F.~J.}\ \bibnamefont {Low}},\ }\bibfield  {title} {\bibinfo {title} {Low-{Temperature} {Germanium} {Bolometer}},\ }\href {https://doi.org/10.1364/JOSA.51.001300} {\bibfield  {journal} {\bibinfo  {journal} {J. Opt. Soc. Am.}\ }\textbf {\bibinfo {volume} {51}},\ \bibinfo {pages} {1300} (\bibinfo {year} {1961})}\BibitemShut {NoStop}%
\bibitem [{\citenamefont {Richards}(1994)}]{richards_bolometers_1994}%
  \BibitemOpen
  \bibfield  {author} {\bibinfo {author} {\bibfnamefont {P.~L.}\ \bibnamefont {Richards}},\ }\bibfield  {title} {\bibinfo {title} {Bolometers for infrared and millimeter waves},\ }\href {https://doi.org/10.1063/1.357128} {\bibfield  {journal} {\bibinfo  {journal} {Journal of Applied Physics}\ }\textbf {\bibinfo {volume} {76}},\ \bibinfo {pages} {1} (\bibinfo {year} {1994})}\BibitemShut {NoStop}%
\bibitem [{\citenamefont {Kokkoniemi}\ \emph {et~al.}(2019)\citenamefont {Kokkoniemi}, \citenamefont {Govenius}, \citenamefont {Vesterinen}, \citenamefont {Lake}, \citenamefont {Gunyhó}, \citenamefont {Tan}, \citenamefont {Simbierowicz}, \citenamefont {Grönberg}, \citenamefont {Lehtinen}, \citenamefont {Prunnila}, \citenamefont {Hassel}, \citenamefont {Lamminen}, \citenamefont {Saira},\ and\ \citenamefont {Möttönen}}]{kokkoniemi_nanobolometer_2019}%
  \BibitemOpen
  \bibfield  {author} {\bibinfo {author} {\bibfnamefont {R.}~\bibnamefont {Kokkoniemi}}, \bibinfo {author} {\bibfnamefont {J.}~\bibnamefont {Govenius}}, \bibinfo {author} {\bibfnamefont {V.}~\bibnamefont {Vesterinen}}, \bibinfo {author} {\bibfnamefont {R.~E.}\ \bibnamefont {Lake}}, \bibinfo {author} {\bibfnamefont {A.~M.}\ \bibnamefont {Gunyhó}}, \bibinfo {author} {\bibfnamefont {K.~Y.}\ \bibnamefont {Tan}}, \bibinfo {author} {\bibfnamefont {S.}~\bibnamefont {Simbierowicz}}, \bibinfo {author} {\bibfnamefont {L.}~\bibnamefont {Grönberg}}, \bibinfo {author} {\bibfnamefont {J.}~\bibnamefont {Lehtinen}}, \bibinfo {author} {\bibfnamefont {M.}~\bibnamefont {Prunnila}}, \bibinfo {author} {\bibfnamefont {J.}~\bibnamefont {Hassel}}, \bibinfo {author} {\bibfnamefont {A.}~\bibnamefont {Lamminen}}, \bibinfo {author} {\bibfnamefont {O.-P.}\ \bibnamefont {Saira}},\ and\ \bibinfo {author} {\bibfnamefont {M.}~\bibnamefont {Möttönen}},\ }\bibfield  {title} {\bibinfo {title} {Nanobolometer with ultralow noise equivalent
  power},\ }\href {https://doi.org/10.1038/s42005-019-0225-6} {\bibfield  {journal} {\bibinfo  {journal} {Commun. Phys.}\ }\textbf {\bibinfo {volume} {2}},\ \bibinfo {pages} {1} (\bibinfo {year} {2019})}\BibitemShut {NoStop}%
\bibitem [{\citenamefont {Jin}\ \emph {et~al.}(2015)\citenamefont {Jin}, \citenamefont {Kamal}, \citenamefont {Sears}, \citenamefont {Gudmundsen}, \citenamefont {Hover}, \citenamefont {Miloshi}, \citenamefont {Slattery}, \citenamefont {Yan}, \citenamefont {Yoder}, \citenamefont {Orlando}, \citenamefont {Gustavsson},\ and\ \citenamefont {Oliver}}]{jin_thermal_2015}%
  \BibitemOpen
  \bibfield  {author} {\bibinfo {author} {\bibfnamefont {X.~Y.}\ \bibnamefont {Jin}}, \bibinfo {author} {\bibfnamefont {A.}~\bibnamefont {Kamal}}, \bibinfo {author} {\bibfnamefont {A.~P.}\ \bibnamefont {Sears}}, \bibinfo {author} {\bibfnamefont {T.}~\bibnamefont {Gudmundsen}}, \bibinfo {author} {\bibfnamefont {D.}~\bibnamefont {Hover}}, \bibinfo {author} {\bibfnamefont {J.}~\bibnamefont {Miloshi}}, \bibinfo {author} {\bibfnamefont {R.}~\bibnamefont {Slattery}}, \bibinfo {author} {\bibfnamefont {F.}~\bibnamefont {Yan}}, \bibinfo {author} {\bibfnamefont {J.}~\bibnamefont {Yoder}}, \bibinfo {author} {\bibfnamefont {T.~P.}\ \bibnamefont {Orlando}}, \bibinfo {author} {\bibfnamefont {S.}~\bibnamefont {Gustavsson}},\ and\ \bibinfo {author} {\bibfnamefont {W.~D.}\ \bibnamefont {Oliver}},\ }\bibfield  {title} {\bibinfo {title} {Thermal and {Residual} {Excited}-{State} {Population} in a {3D} {Transmon} {Qubit}},\ }\href {https://doi.org/10.1103/PhysRevLett.114.240501} {\bibfield  {journal} {\bibinfo  {journal} {Phys.
  Rev. Lett.}\ }\textbf {\bibinfo {volume} {114}},\ \bibinfo {pages} {240501} (\bibinfo {year} {2015})}\BibitemShut {NoStop}%
\bibitem [{\citenamefont {Kulikov}\ \emph {et~al.}(2020)\citenamefont {Kulikov}, \citenamefont {Navarathna},\ and\ \citenamefont {Fedorov}}]{kulikov_measuring_2020}%
  \BibitemOpen
  \bibfield  {author} {\bibinfo {author} {\bibfnamefont {A.}~\bibnamefont {Kulikov}}, \bibinfo {author} {\bibfnamefont {R.}~\bibnamefont {Navarathna}},\ and\ \bibinfo {author} {\bibfnamefont {A.}~\bibnamefont {Fedorov}},\ }\bibfield  {title} {\bibinfo {title} {Measuring {Effective} {Temperatures} of {Qubits} {Using} {Correlations}},\ }\href {https://doi.org/10.1103/PhysRevLett.124.240501} {\bibfield  {journal} {\bibinfo  {journal} {Phys. Rev. Lett.}\ }\textbf {\bibinfo {volume} {124}},\ \bibinfo {pages} {240501} (\bibinfo {year} {2020})}\BibitemShut {NoStop}%
\bibitem [{\citenamefont {Sultanov}\ \emph {et~al.}(2021)\citenamefont {Sultanov}, \citenamefont {Kuzmanović}, \citenamefont {Lebedev},\ and\ \citenamefont {Paraoanu}}]{sultanov_protocol_2021}%
  \BibitemOpen
  \bibfield  {author} {\bibinfo {author} {\bibfnamefont {A.}~\bibnamefont {Sultanov}}, \bibinfo {author} {\bibfnamefont {M.}~\bibnamefont {Kuzmanović}}, \bibinfo {author} {\bibfnamefont {A.~V.}\ \bibnamefont {Lebedev}},\ and\ \bibinfo {author} {\bibfnamefont {G.~S.}\ \bibnamefont {Paraoanu}},\ }\bibfield  {title} {\bibinfo {title} {Protocol for temperature sensing using a three-level transmon circuit},\ }\href {https://doi.org/10.1063/5.0065224} {\bibfield  {journal} {\bibinfo  {journal} {Applied Physics Letters}\ }\textbf {\bibinfo {volume} {119}},\ \bibinfo {pages} {144002} (\bibinfo {year} {2021})}\BibitemShut {NoStop}%
\bibitem [{\citenamefont {Lvov}\ \emph {et~al.}(2025)\citenamefont {Lvov}, \citenamefont {Lemziakov}, \citenamefont {Ankerhold}, \citenamefont {Peltonen},\ and\ \citenamefont {Pekola}}]{lvov_thermometry_2025}%
  \BibitemOpen
  \bibfield  {author} {\bibinfo {author} {\bibfnamefont {D.~S.}\ \bibnamefont {Lvov}}, \bibinfo {author} {\bibfnamefont {S.~A.}\ \bibnamefont {Lemziakov}}, \bibinfo {author} {\bibfnamefont {E.}~\bibnamefont {Ankerhold}}, \bibinfo {author} {\bibfnamefont {J.~T.}\ \bibnamefont {Peltonen}},\ and\ \bibinfo {author} {\bibfnamefont {J.~P.}\ \bibnamefont {Pekola}},\ }\bibfield  {title} {\bibinfo {title} {Thermometry based on a superconducting qubit},\ }\href {https://doi.org/10.1103/PhysRevApplied.23.054079} {\bibfield  {journal} {\bibinfo  {journal} {Phys. Rev. Appl.}\ }\textbf {\bibinfo {volume} {23}},\ \bibinfo {pages} {054079} (\bibinfo {year} {2025})}\BibitemShut {NoStop}%
\bibitem [{\citenamefont {Paik}\ \emph {et~al.}(2011)\citenamefont {Paik}, \citenamefont {Schuster}, \citenamefont {Bishop}, \citenamefont {Kirchmair}, \citenamefont {Catelani}, \citenamefont {Sears}, \citenamefont {Johnson}, \citenamefont {Reagor}, \citenamefont {Frunzio}, \citenamefont {Glazman}, \citenamefont {Girvin}, \citenamefont {Devoret},\ and\ \citenamefont {Schoelkopf}}]{paik_observation_2011}%
  \BibitemOpen
  \bibfield  {author} {\bibinfo {author} {\bibfnamefont {H.}~\bibnamefont {Paik}}, \bibinfo {author} {\bibfnamefont {D.~I.}\ \bibnamefont {Schuster}}, \bibinfo {author} {\bibfnamefont {L.~S.}\ \bibnamefont {Bishop}}, \bibinfo {author} {\bibfnamefont {G.}~\bibnamefont {Kirchmair}}, \bibinfo {author} {\bibfnamefont {G.}~\bibnamefont {Catelani}}, \bibinfo {author} {\bibfnamefont {A.~P.}\ \bibnamefont {Sears}}, \bibinfo {author} {\bibfnamefont {B.~R.}\ \bibnamefont {Johnson}}, \bibinfo {author} {\bibfnamefont {M.~J.}\ \bibnamefont {Reagor}}, \bibinfo {author} {\bibfnamefont {L.}~\bibnamefont {Frunzio}}, \bibinfo {author} {\bibfnamefont {L.~I.}\ \bibnamefont {Glazman}}, \bibinfo {author} {\bibfnamefont {S.~M.}\ \bibnamefont {Girvin}}, \bibinfo {author} {\bibfnamefont {M.~H.}\ \bibnamefont {Devoret}},\ and\ \bibinfo {author} {\bibfnamefont {R.~J.}\ \bibnamefont {Schoelkopf}},\ }\bibfield  {title} {\bibinfo {title} {Observation of {High} {Coherence} in {Josephson} {Junction} {Qubits} {Measured} in a
  {Three}-{Dimensional} {Circuit} {QED} {Architecture}},\ }\href {https://doi.org/10.1103/PhysRevLett.107.240501} {\bibfield  {journal} {\bibinfo  {journal} {Phys. Rev. Lett.}\ }\textbf {\bibinfo {volume} {107}},\ \bibinfo {pages} {240501} (\bibinfo {year} {2011})}\BibitemShut {NoStop}%
\bibitem [{\citenamefont {Burnett}\ \emph {et~al.}(2014)\citenamefont {Burnett}, \citenamefont {Faoro}, \citenamefont {Wisby}, \citenamefont {Gurtovoi}, \citenamefont {Chernykh}, \citenamefont {Mikhailov}, \citenamefont {Tulin}, \citenamefont {Shaikhaidarov}, \citenamefont {Antonov}, \citenamefont {Meeson}, \citenamefont {Tzalenchuk},\ and\ \citenamefont {Lindström}}]{burnett_evidence_2014}%
  \BibitemOpen
  \bibfield  {author} {\bibinfo {author} {\bibfnamefont {J.}~\bibnamefont {Burnett}}, \bibinfo {author} {\bibfnamefont {L.}~\bibnamefont {Faoro}}, \bibinfo {author} {\bibfnamefont {I.}~\bibnamefont {Wisby}}, \bibinfo {author} {\bibfnamefont {V.~L.}\ \bibnamefont {Gurtovoi}}, \bibinfo {author} {\bibfnamefont {A.~V.}\ \bibnamefont {Chernykh}}, \bibinfo {author} {\bibfnamefont {G.~M.}\ \bibnamefont {Mikhailov}}, \bibinfo {author} {\bibfnamefont {V.~A.}\ \bibnamefont {Tulin}}, \bibinfo {author} {\bibfnamefont {R.}~\bibnamefont {Shaikhaidarov}}, \bibinfo {author} {\bibfnamefont {V.}~\bibnamefont {Antonov}}, \bibinfo {author} {\bibfnamefont {P.~J.}\ \bibnamefont {Meeson}}, \bibinfo {author} {\bibfnamefont {A.~Y.}\ \bibnamefont {Tzalenchuk}},\ and\ \bibinfo {author} {\bibfnamefont {T.}~\bibnamefont {Lindström}},\ }\bibfield  {title} {\bibinfo {title} {Evidence for interacting two-level systems from the 1/f noise of a superconducting resonator},\ }\href {https://doi.org/10.1038/ncomms5119} {\bibfield  {journal}
  {\bibinfo  {journal} {Nat. Commun.}\ }\textbf {\bibinfo {volume} {5}},\ \bibinfo {pages} {4119} (\bibinfo {year} {2014})}\BibitemShut {NoStop}%
\bibitem [{\citenamefont {de~Graaf}\ \emph {et~al.}(2018)\citenamefont {de~Graaf}, \citenamefont {Faoro}, \citenamefont {Burnett}, \citenamefont {Adamyan}, \citenamefont {Tzalenchuk}, \citenamefont {Kubatkin}, \citenamefont {Lindström},\ and\ \citenamefont {Danilov}}]{de_graaf_suppression_2018}%
  \BibitemOpen
  \bibfield  {author} {\bibinfo {author} {\bibfnamefont {S.~E.}\ \bibnamefont {de~Graaf}}, \bibinfo {author} {\bibfnamefont {L.}~\bibnamefont {Faoro}}, \bibinfo {author} {\bibfnamefont {J.}~\bibnamefont {Burnett}}, \bibinfo {author} {\bibfnamefont {A.~A.}\ \bibnamefont {Adamyan}}, \bibinfo {author} {\bibfnamefont {A.~Y.}\ \bibnamefont {Tzalenchuk}}, \bibinfo {author} {\bibfnamefont {S.~E.}\ \bibnamefont {Kubatkin}}, \bibinfo {author} {\bibfnamefont {T.}~\bibnamefont {Lindström}},\ and\ \bibinfo {author} {\bibfnamefont {A.~V.}\ \bibnamefont {Danilov}},\ }\bibfield  {title} {\bibinfo {title} {Suppression of low-frequency charge noise in superconducting resonators by surface spin desorption},\ }\href {https://doi.org/10.1038/s41467-018-03577-2} {\bibfield  {journal} {\bibinfo  {journal} {Nat. Commun.}\ }\textbf {\bibinfo {volume} {9}},\ \bibinfo {pages} {1143} (\bibinfo {year} {2018})}\BibitemShut {NoStop}%
\bibitem [{\citenamefont {Lucas}\ \emph {et~al.}(2023)\citenamefont {Lucas}, \citenamefont {Danilov}, \citenamefont {Levitin}, \citenamefont {Jayaraman}, \citenamefont {Casey}, \citenamefont {Faoro}, \citenamefont {Tzalenchuk}, \citenamefont {Kubatkin}, \citenamefont {Saunders},\ and\ \citenamefont {de~Graaf}}]{lucas_quantum_2023}%
  \BibitemOpen
  \bibfield  {author} {\bibinfo {author} {\bibfnamefont {M.}~\bibnamefont {Lucas}}, \bibinfo {author} {\bibfnamefont {A.~V.}\ \bibnamefont {Danilov}}, \bibinfo {author} {\bibfnamefont {L.~V.}\ \bibnamefont {Levitin}}, \bibinfo {author} {\bibfnamefont {A.}~\bibnamefont {Jayaraman}}, \bibinfo {author} {\bibfnamefont {A.~J.}\ \bibnamefont {Casey}}, \bibinfo {author} {\bibfnamefont {L.}~\bibnamefont {Faoro}}, \bibinfo {author} {\bibfnamefont {A.~Y.}\ \bibnamefont {Tzalenchuk}}, \bibinfo {author} {\bibfnamefont {S.~E.}\ \bibnamefont {Kubatkin}}, \bibinfo {author} {\bibfnamefont {J.}~\bibnamefont {Saunders}},\ and\ \bibinfo {author} {\bibfnamefont {S.~E.}\ \bibnamefont {de~Graaf}},\ }\bibfield  {title} {\bibinfo {title} {Quantum bath suppression in a superconducting circuit by immersion cooling},\ }\href {https://doi.org/10.1038/s41467-023-39249-z} {\bibfield  {journal} {\bibinfo  {journal} {Nat. Commun.}\ }\textbf {\bibinfo {volume} {14}},\ \bibinfo {pages} {3522} (\bibinfo {year} {2023})}\BibitemShut {NoStop}%
\bibitem [{\citenamefont {de~Graaf}\ \emph {et~al.}(2017)\citenamefont {de~Graaf}, \citenamefont {Adamyan}, \citenamefont {Lindström}, \citenamefont {Erts}, \citenamefont {Kubatkin}, \citenamefont {Tzalenchuk},\ and\ \citenamefont {Danilov}}]{de_graaf_direct_2017}%
  \BibitemOpen
  \bibfield  {author} {\bibinfo {author} {\bibfnamefont {S.~E.}\ \bibnamefont {de~Graaf}}, \bibinfo {author} {\bibfnamefont {A.~A.}\ \bibnamefont {Adamyan}}, \bibinfo {author} {\bibfnamefont {T.}~\bibnamefont {Lindström}}, \bibinfo {author} {\bibfnamefont {D.}~\bibnamefont {Erts}}, \bibinfo {author} {\bibfnamefont {S.~E.}\ \bibnamefont {Kubatkin}}, \bibinfo {author} {\bibfnamefont {A.~Y.}\ \bibnamefont {Tzalenchuk}},\ and\ \bibinfo {author} {\bibfnamefont {A.~V.}\ \bibnamefont {Danilov}},\ }\bibfield  {title} {\bibinfo {title} {Direct {Identification} of {Dilute} {Surface} {Spins} on {Al2O3}: {Origin} of {Flux} {Noise} in {Quantum} {Circuits}},\ }\href {https://doi.org/10.1103/PhysRevLett.118.057703} {\bibfield  {journal} {\bibinfo  {journal} {Phys. Rev. Lett.}\ }\textbf {\bibinfo {volume} {118}},\ \bibinfo {pages} {057703} (\bibinfo {year} {2017})}\BibitemShut {NoStop}%
\bibitem [{\citenamefont {Quintana}\ \emph {et~al.}(2017)\citenamefont {Quintana}, \citenamefont {Chen}, \citenamefont {Sank}, \citenamefont {Petukhov}, \citenamefont {White}, \citenamefont {Kafri}, \citenamefont {Chiaro}, \citenamefont {Megrant}, \citenamefont {Barends}, \citenamefont {Campbell}, \citenamefont {Chen}, \citenamefont {Dunsworth}, \citenamefont {Fowler}, \citenamefont {Graff}, \citenamefont {Jeffrey}, \citenamefont {Kelly}, \citenamefont {Lucero}, \citenamefont {Mutus}, \citenamefont {Neeley}, \citenamefont {Neill}, \citenamefont {O’Malley}, \citenamefont {Roushan}, \citenamefont {Shabani}, \citenamefont {Smelyanskiy}, \citenamefont {Vainsencher}, \citenamefont {Wenner}, \citenamefont {Neven},\ and\ \citenamefont {Martinis}}]{quintana_observation_2017}%
  \BibitemOpen
  \bibfield  {author} {\bibinfo {author} {\bibfnamefont {C.~M.}\ \bibnamefont {Quintana}}, \bibinfo {author} {\bibfnamefont {Y.}~\bibnamefont {Chen}}, \bibinfo {author} {\bibfnamefont {D.}~\bibnamefont {Sank}}, \bibinfo {author} {\bibfnamefont {A.~G.}\ \bibnamefont {Petukhov}}, \bibinfo {author} {\bibfnamefont {T.~C.}\ \bibnamefont {White}}, \bibinfo {author} {\bibfnamefont {D.}~\bibnamefont {Kafri}}, \bibinfo {author} {\bibfnamefont {B.}~\bibnamefont {Chiaro}}, \bibinfo {author} {\bibfnamefont {A.}~\bibnamefont {Megrant}}, \bibinfo {author} {\bibfnamefont {R.}~\bibnamefont {Barends}}, \bibinfo {author} {\bibfnamefont {B.}~\bibnamefont {Campbell}}, \bibinfo {author} {\bibfnamefont {Z.}~\bibnamefont {Chen}}, \bibinfo {author} {\bibfnamefont {A.}~\bibnamefont {Dunsworth}}, \bibinfo {author} {\bibfnamefont {A.~G.}\ \bibnamefont {Fowler}}, \bibinfo {author} {\bibfnamefont {R.}~\bibnamefont {Graff}}, \bibinfo {author} {\bibfnamefont {E.}~\bibnamefont {Jeffrey}}, \bibinfo {author} {\bibfnamefont {J.}~\bibnamefont
  {Kelly}}, \bibinfo {author} {\bibfnamefont {E.}~\bibnamefont {Lucero}}, \bibinfo {author} {\bibfnamefont {J.~Y.}\ \bibnamefont {Mutus}}, \bibinfo {author} {\bibfnamefont {M.}~\bibnamefont {Neeley}}, \bibinfo {author} {\bibfnamefont {C.}~\bibnamefont {Neill}}, \bibinfo {author} {\bibfnamefont {P.~J.~J.}\ \bibnamefont {O’Malley}}, \bibinfo {author} {\bibfnamefont {P.}~\bibnamefont {Roushan}}, \bibinfo {author} {\bibfnamefont {A.}~\bibnamefont {Shabani}}, \bibinfo {author} {\bibfnamefont {V.~N.}\ \bibnamefont {Smelyanskiy}}, \bibinfo {author} {\bibfnamefont {A.}~\bibnamefont {Vainsencher}}, \bibinfo {author} {\bibfnamefont {J.}~\bibnamefont {Wenner}}, \bibinfo {author} {\bibfnamefont {H.}~\bibnamefont {Neven}},\ and\ \bibinfo {author} {\bibfnamefont {J.~M.}\ \bibnamefont {Martinis}},\ }\bibfield  {title} {\bibinfo {title} {Observation of {Classical}-{Quantum} {Crossover} of 1/f {Flux} {Noise} and {Its} {Paramagnetic} {Temperature} {Dependence}},\ }\href {https://doi.org/10.1103/PhysRevLett.118.057702}
  {\bibfield  {journal} {\bibinfo  {journal} {Phys. Rev. Lett.}\ }\textbf {\bibinfo {volume} {118}},\ \bibinfo {pages} {057702} (\bibinfo {year} {2017})}\BibitemShut {NoStop}%
\bibitem [{\citenamefont {Satrya}\ \emph {et~al.}(2025)\citenamefont {Satrya}, \citenamefont {Chang}, \citenamefont {Strelnikov}, \citenamefont {Upadhyay}, \citenamefont {Mäkinen}, \citenamefont {Peltonen}, \citenamefont {Karimi},\ and\ \citenamefont {Pekola}}]{satrya_thermal_2025}%
  \BibitemOpen
  \bibfield  {author} {\bibinfo {author} {\bibfnamefont {C.~D.}\ \bibnamefont {Satrya}}, \bibinfo {author} {\bibfnamefont {Y.-C.}\ \bibnamefont {Chang}}, \bibinfo {author} {\bibfnamefont {A.~S.}\ \bibnamefont {Strelnikov}}, \bibinfo {author} {\bibfnamefont {R.}~\bibnamefont {Upadhyay}}, \bibinfo {author} {\bibfnamefont {I.~K.}\ \bibnamefont {Mäkinen}}, \bibinfo {author} {\bibfnamefont {J.~T.}\ \bibnamefont {Peltonen}}, \bibinfo {author} {\bibfnamefont {B.}~\bibnamefont {Karimi}},\ and\ \bibinfo {author} {\bibfnamefont {J.~P.}\ \bibnamefont {Pekola}},\ }\bibfield  {title} {\bibinfo {title} {Thermal spectrometer for superconducting circuits},\ }\href {https://doi.org/10.1038/s41467-025-58919-8} {\bibfield  {journal} {\bibinfo  {journal} {Nat. Commun.}\ }\textbf {\bibinfo {volume} {16}},\ \bibinfo {pages} {4435} (\bibinfo {year} {2025})}\BibitemShut {NoStop}%
\bibitem [{\citenamefont {Castro~Neto}\ \emph {et~al.}(2009)\citenamefont {Castro~Neto}, \citenamefont {Guinea}, \citenamefont {Peres}, \citenamefont {Novoselov},\ and\ \citenamefont {Geim}}]{castro_neto_electronic_2009}%
  \BibitemOpen
  \bibfield  {author} {\bibinfo {author} {\bibfnamefont {A.~H.}\ \bibnamefont {Castro~Neto}}, \bibinfo {author} {\bibfnamefont {F.}~\bibnamefont {Guinea}}, \bibinfo {author} {\bibfnamefont {N.~M.~R.}\ \bibnamefont {Peres}}, \bibinfo {author} {\bibfnamefont {K.~S.}\ \bibnamefont {Novoselov}},\ and\ \bibinfo {author} {\bibfnamefont {A.~K.}\ \bibnamefont {Geim}},\ }\bibfield  {title} {\bibinfo {title} {The electronic properties of graphene},\ }\href {https://doi.org/10.1103/RevModPhys.81.109} {\bibfield  {journal} {\bibinfo  {journal} {Rev. Mod. Phys.}\ }\textbf {\bibinfo {volume} {81}},\ \bibinfo {pages} {109} (\bibinfo {year} {2009})}\BibitemShut {NoStop}%
\bibitem [{\citenamefont {Das~Sarma}\ \emph {et~al.}(2011)\citenamefont {Das~Sarma}, \citenamefont {Adam}, \citenamefont {Hwang},\ and\ \citenamefont {Rossi}}]{das_sarma_electronic_2011}%
  \BibitemOpen
  \bibfield  {author} {\bibinfo {author} {\bibfnamefont {S.}~\bibnamefont {Das~Sarma}}, \bibinfo {author} {\bibfnamefont {S.}~\bibnamefont {Adam}}, \bibinfo {author} {\bibfnamefont {E.~H.}\ \bibnamefont {Hwang}},\ and\ \bibinfo {author} {\bibfnamefont {E.}~\bibnamefont {Rossi}},\ }\bibfield  {title} {\bibinfo {title} {Electronic transport in two-dimensional graphene},\ }\href {https://doi.org/10.1103/RevModPhys.83.407} {\bibfield  {journal} {\bibinfo  {journal} {Rev. Mod. Phys.}\ }\textbf {\bibinfo {volume} {83}},\ \bibinfo {pages} {407} (\bibinfo {year} {2011})}\BibitemShut {NoStop}%
\bibitem [{\citenamefont {Moseley}\ \emph {et~al.}(1984)\citenamefont {Moseley}, \citenamefont {Mather},\ and\ \citenamefont {McCammon}}]{moseley_thermal_1984}%
  \BibitemOpen
  \bibfield  {author} {\bibinfo {author} {\bibfnamefont {S.~H.}\ \bibnamefont {Moseley}}, \bibinfo {author} {\bibfnamefont {J.~C.}\ \bibnamefont {Mather}},\ and\ \bibinfo {author} {\bibfnamefont {D.}~\bibnamefont {McCammon}},\ }\bibfield  {title} {\bibinfo {title} {Thermal detectors as x‐ray spectrometers},\ }\href {https://doi.org/10.1063/1.334129} {\bibfield  {journal} {\bibinfo  {journal} {Journal of Applied Physics}\ }\textbf {\bibinfo {volume} {56}},\ \bibinfo {pages} {1257} (\bibinfo {year} {1984})}\BibitemShut {NoStop}%
\bibitem [{\citenamefont {Lee}\ \emph {et~al.}(2020)\citenamefont {Lee}, \citenamefont {Efetov}, \citenamefont {Jung}, \citenamefont {Ranzani}, \citenamefont {Walsh}, \citenamefont {Ohki}, \citenamefont {Taniguchi}, \citenamefont {Watanabe}, \citenamefont {Kim}, \citenamefont {Englund},\ and\ \citenamefont {Fong}}]{lee_graphene-based_2020}%
  \BibitemOpen
  \bibfield  {author} {\bibinfo {author} {\bibfnamefont {G.-H.}\ \bibnamefont {Lee}}, \bibinfo {author} {\bibfnamefont {D.~K.}\ \bibnamefont {Efetov}}, \bibinfo {author} {\bibfnamefont {W.}~\bibnamefont {Jung}}, \bibinfo {author} {\bibfnamefont {L.}~\bibnamefont {Ranzani}}, \bibinfo {author} {\bibfnamefont {E.~D.}\ \bibnamefont {Walsh}}, \bibinfo {author} {\bibfnamefont {T.~A.}\ \bibnamefont {Ohki}}, \bibinfo {author} {\bibfnamefont {T.}~\bibnamefont {Taniguchi}}, \bibinfo {author} {\bibfnamefont {K.}~\bibnamefont {Watanabe}}, \bibinfo {author} {\bibfnamefont {P.}~\bibnamefont {Kim}}, \bibinfo {author} {\bibfnamefont {D.}~\bibnamefont {Englund}},\ and\ \bibinfo {author} {\bibfnamefont {K.~C.}\ \bibnamefont {Fong}},\ }\bibfield  {title} {\bibinfo {title} {Graphene-based {Josephson} junction microwave bolometer},\ }\href {https://doi.org/10.1038/s41586-020-2752-4} {\bibfield  {journal} {\bibinfo  {journal} {Nature}\ }\textbf {\bibinfo {volume} {586}},\ \bibinfo {pages} {42} (\bibinfo {year} {2020})}\BibitemShut
  {NoStop}%
\bibitem [{\citenamefont {Pereira}\ \emph {et~al.}(2023)\citenamefont {Pereira}, \citenamefont {García-Ripoll},\ and\ \citenamefont {Ramos}}]{pereira_parallel_2023}%
  \BibitemOpen
  \bibfield  {author} {\bibinfo {author} {\bibfnamefont {L.}~\bibnamefont {Pereira}}, \bibinfo {author} {\bibfnamefont {J.~J.}\ \bibnamefont {García-Ripoll}},\ and\ \bibinfo {author} {\bibfnamefont {T.}~\bibnamefont {Ramos}},\ }\bibfield  {title} {\bibinfo {title} {Parallel tomography of quantum non-demolition measurements in multi-qubit devices},\ }\href {https://doi.org/10.1038/s41534-023-00688-7} {\bibfield  {journal} {\bibinfo  {journal} {npj Quantum Inf}\ }\textbf {\bibinfo {volume} {9}},\ \bibinfo {pages} {1} (\bibinfo {year} {2023})}\BibitemShut {NoStop}%
\bibitem [{\citenamefont {Paolucci}\ \emph {et~al.}(2021)\citenamefont {Paolucci}, \citenamefont {Ligato}, \citenamefont {Germanese}, \citenamefont {Buccheri},\ and\ \citenamefont {Giazotto}}]{paolucci_fully_2021}%
  \BibitemOpen
  \bibfield  {author} {\bibinfo {author} {\bibfnamefont {F.}~\bibnamefont {Paolucci}}, \bibinfo {author} {\bibfnamefont {N.}~\bibnamefont {Ligato}}, \bibinfo {author} {\bibfnamefont {G.}~\bibnamefont {Germanese}}, \bibinfo {author} {\bibfnamefont {V.}~\bibnamefont {Buccheri}},\ and\ \bibinfo {author} {\bibfnamefont {F.}~\bibnamefont {Giazotto}},\ }\bibfield  {title} {\bibinfo {title} {Fully {Superconducting} {Josephson} {Bolometers} for {Gigahertz} {Astronomy}},\ }\href {https://doi.org/10.3390/app11020746} {\bibfield  {journal} {\bibinfo  {journal} {Applied Sciences}\ }\textbf {\bibinfo {volume} {11}},\ \bibinfo {pages} {746} (\bibinfo {year} {2021})}\BibitemShut {NoStop}%
\bibitem [{\citenamefont {Ueda}(2020)}]{ueda_quantum_2020}%
  \BibitemOpen
  \bibfield  {author} {\bibinfo {author} {\bibfnamefont {M.}~\bibnamefont {Ueda}},\ }\bibfield  {title} {\bibinfo {title} {Quantum equilibration, thermalization and prethermalization in ultracold atoms},\ }\href {https://doi.org/10.1038/s42254-020-0237-x} {\bibfield  {journal} {\bibinfo  {journal} {Nat. Rev. Phys.}\ }\textbf {\bibinfo {volume} {2}},\ \bibinfo {pages} {669} (\bibinfo {year} {2020})}\BibitemShut {NoStop}%
\end{thebibliography}%

    \begin{figure*}	
        \includegraphics [width=\textwidth] {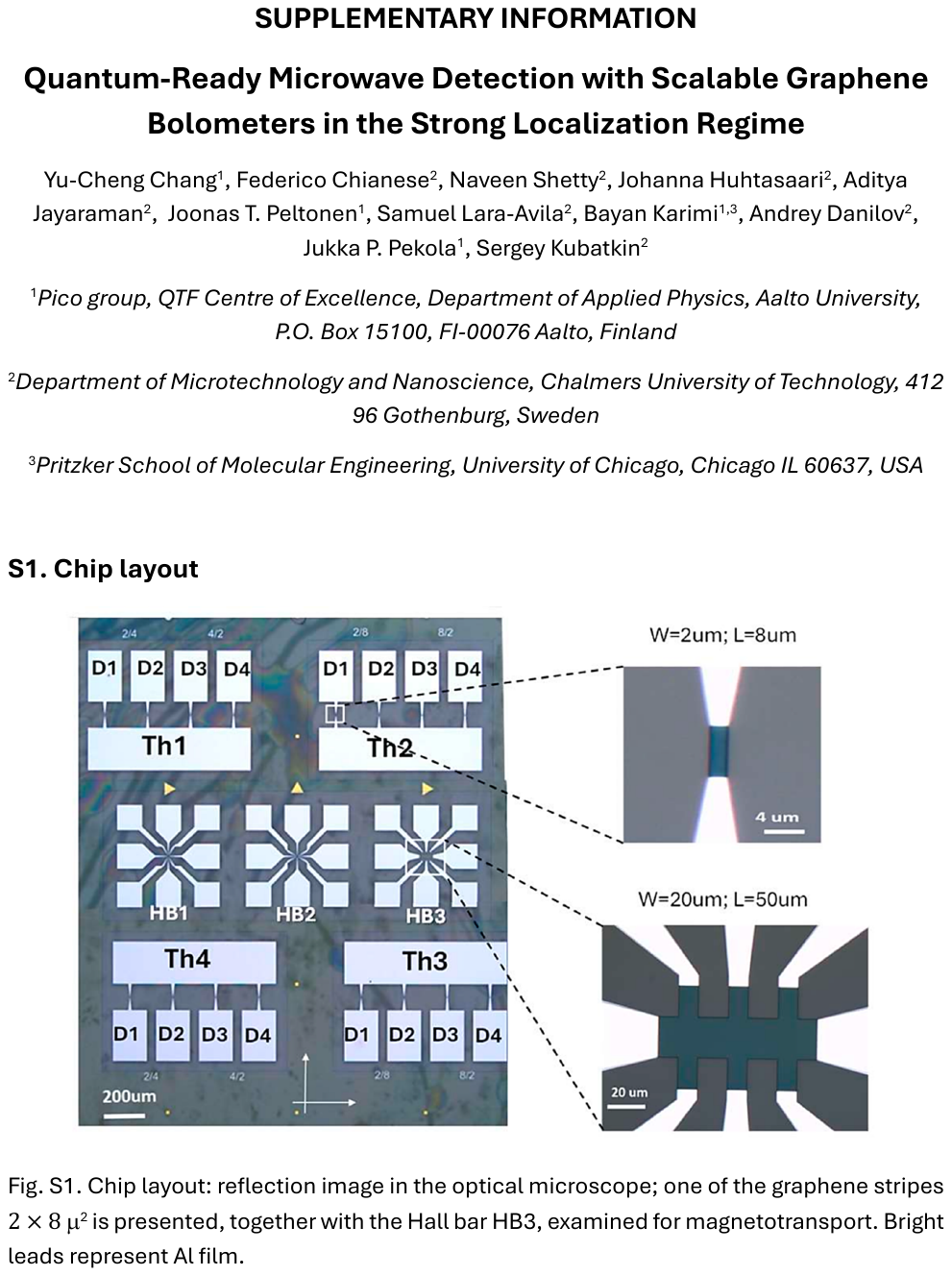}
        \centering
    \end{figure*}    

	\begin{figure*}
		\centering
		\includegraphics [width=\textwidth] {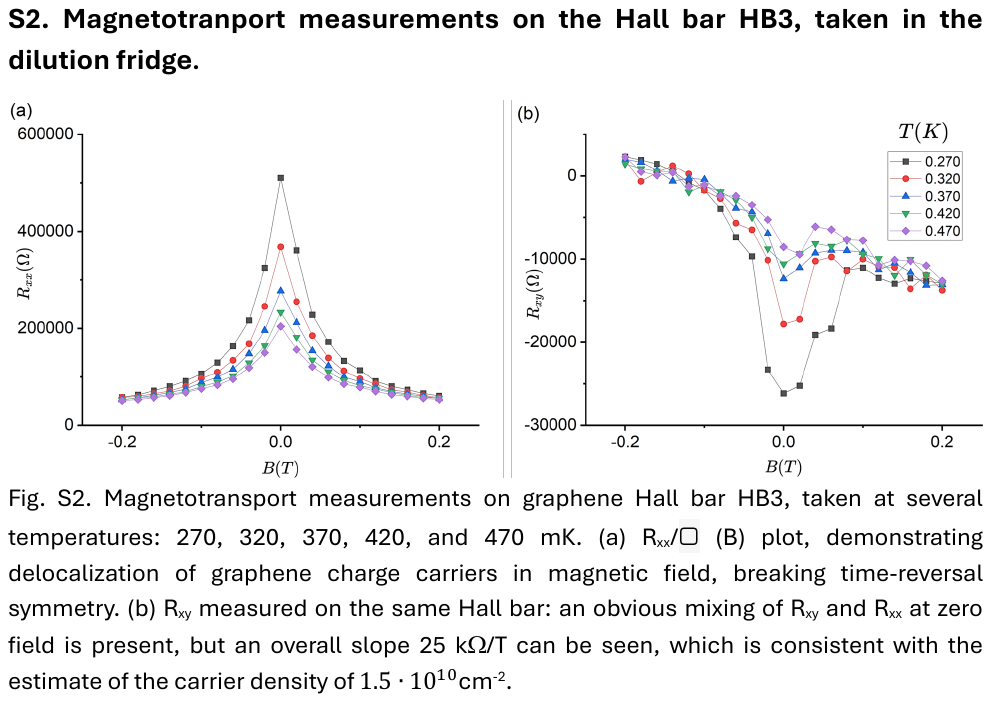}
	\end{figure*}

	\begin{figure*}
		\centering
		\includegraphics [width=\textwidth] {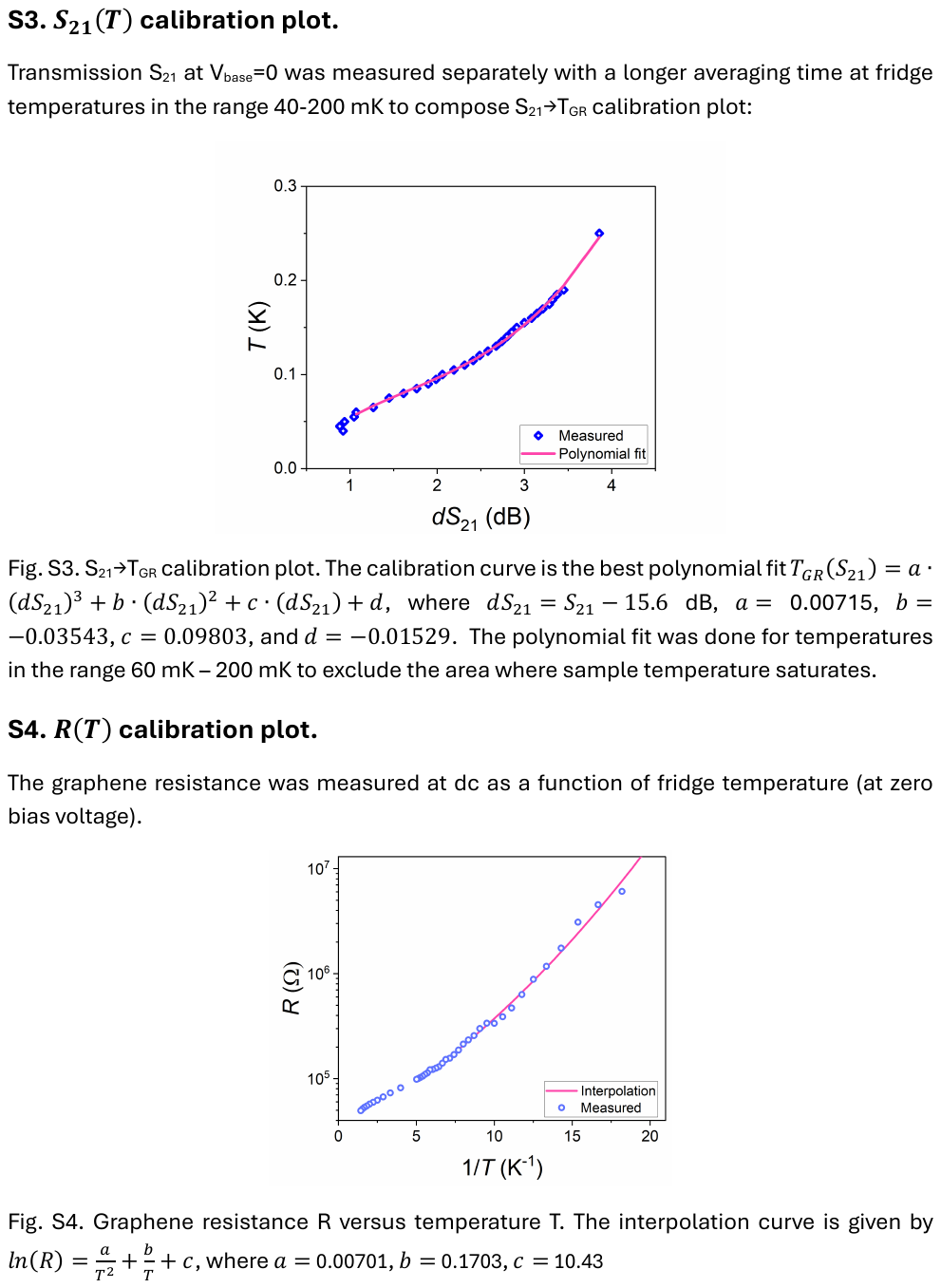}
	\end{figure*}

	\begin{figure*}
		\centering
		\includegraphics [width=\textwidth] {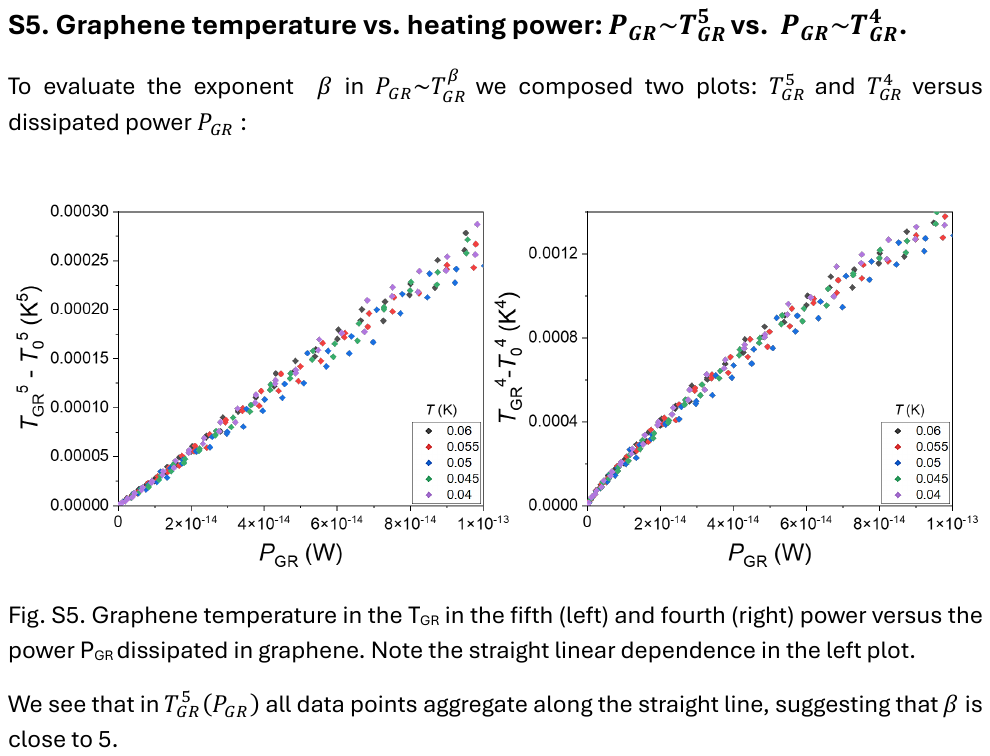}
	\end{figure*}

	\begin{figure*}
		\centering
		\includegraphics [width=\textwidth] {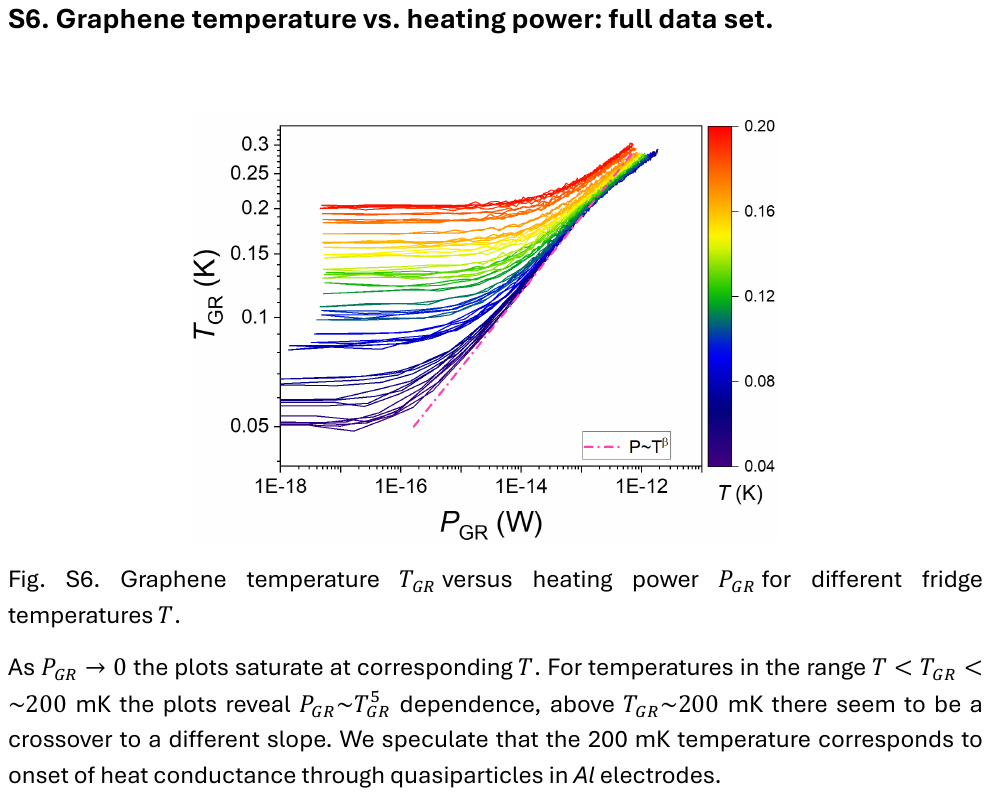}
	\end{figure*}

	\begin{figure*}
		\centering
		\includegraphics [width=\textwidth] {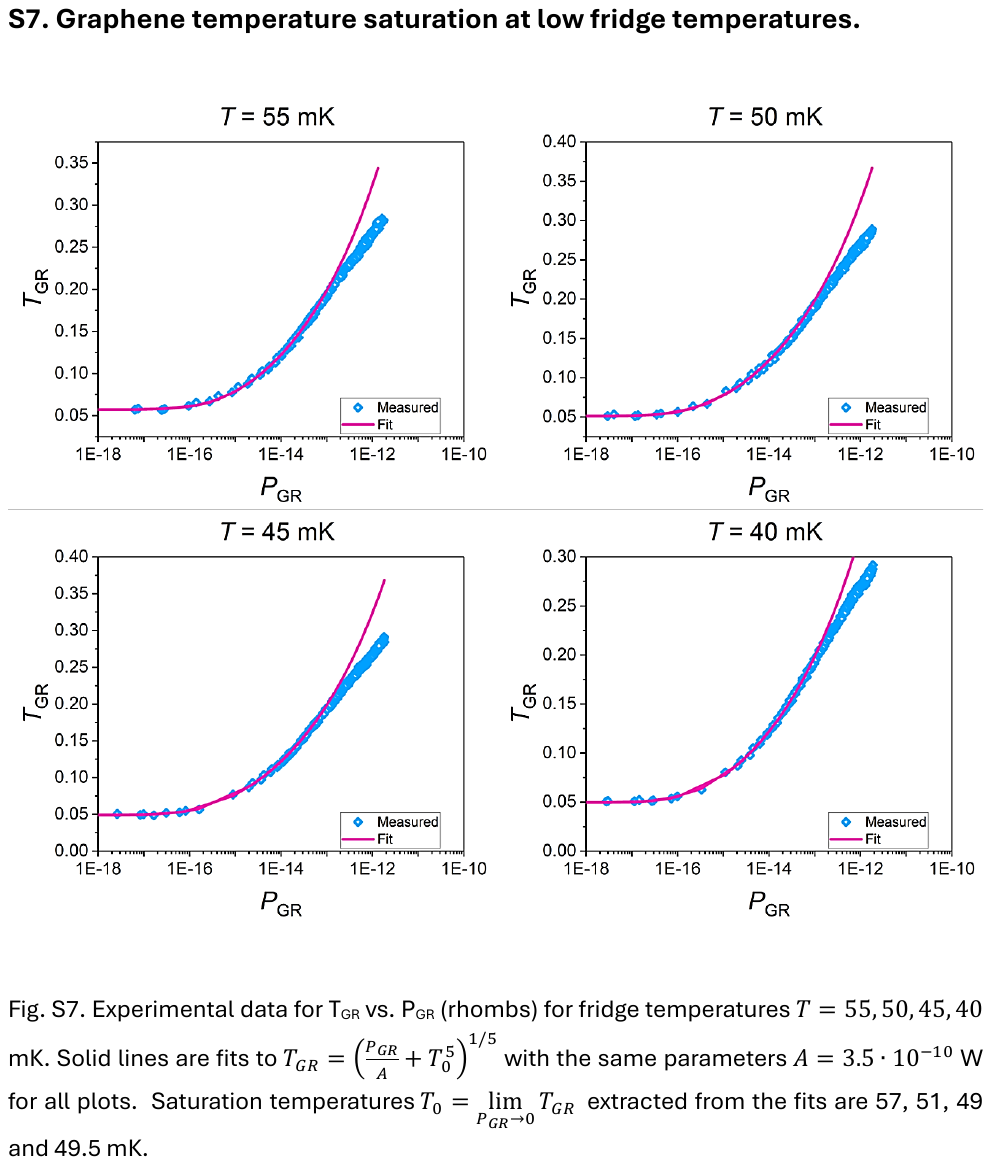}
	\end{figure*}

	\begin{figure*}
		\centering
		\includegraphics [width=\textwidth] {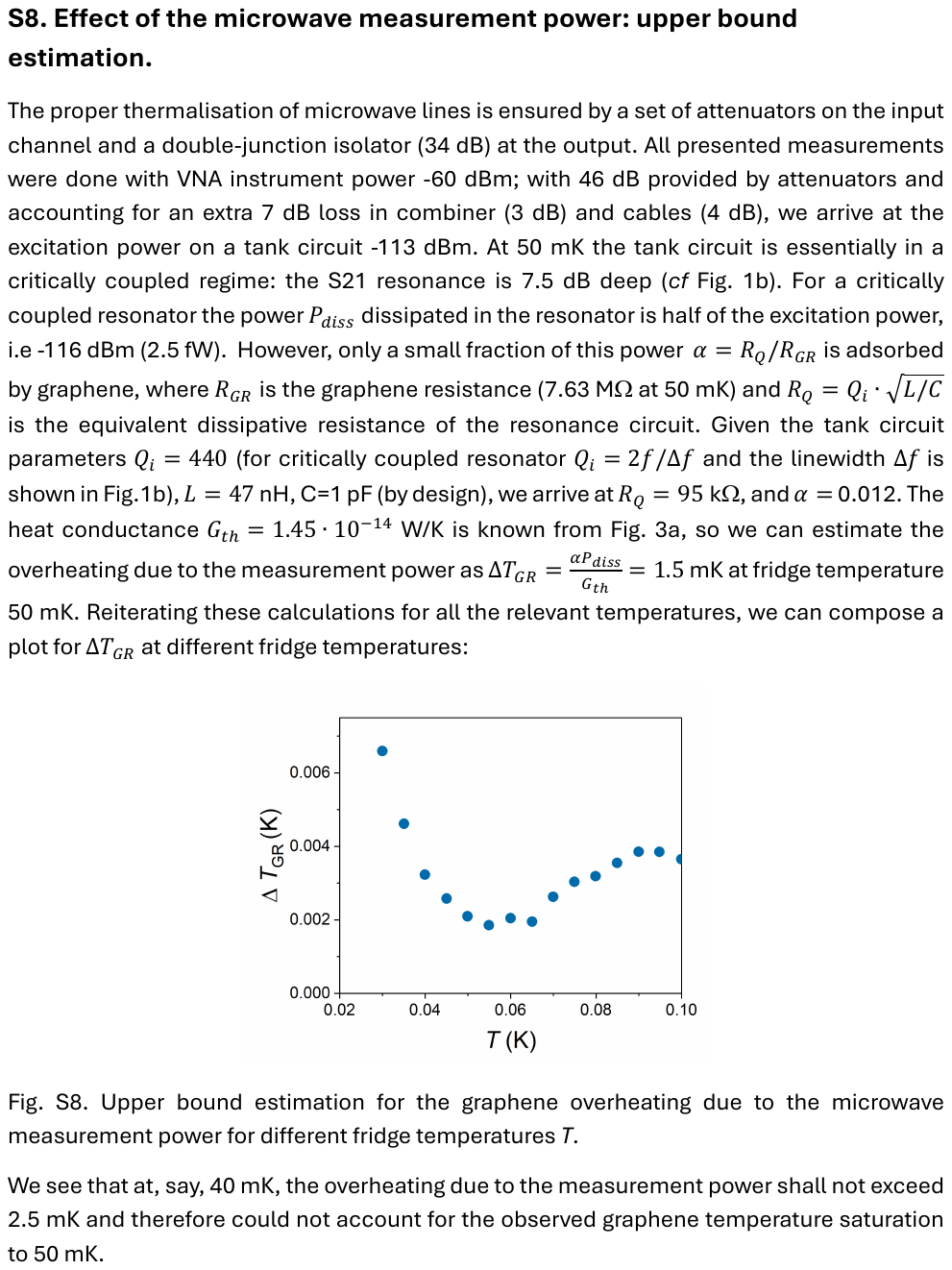}
	\end{figure*}

\end{document}